\documentclass[journal,draftcls,onecolumn,12pt,twoside]{IEEEtran}
\usepackage{algorithm} 
\usepackage{algpseudocode}
\usepackage{graphicx}
\usepackage{amssymb}
\usepackage{amsfonts}
\usepackage[cmex10]{amsmath}
\usepackage{dsfont}
\usepackage{float}
\usepackage{url}
\usepackage{bm}
\usepackage{hhline}
\usepackage{enumerate}
\usepackage{pgf,tikz}
\usepackage{mathrsfs}
\usepackage[framed,numbered,autolinebreaks,useliterate]{mcode}
\usetikzlibrary{arrows}

 \usepackage[colorlinks=true]{hyperref}
\hypersetup{urlcolor=blue, citecolor=blue,linkcolor=purple}

\IEEEoverridecommandlockouts

\newtheorem{Definition}{Definition}

\makeatletter
\let\@@pmod\pmod
\DeclareRobustCommand{\pmod}{\@ifstar\@pmods\@@pmod}
\def\@pmods#1{\mkern4mu({\operator@font mod}\mkern 6mu#1)}
\makeatother

\usepackage{tikz}
\usetikzlibrary{shapes,arrows}
\usepackage{verbatim}
\usetikzlibrary{arrows}

\tikzstyle{startstop} = [rectangle, rounded corners, minimum width=1.5cm, minimum height=0.65cm,text centered, draw=black, fill=blue!3]
\tikzstyle{io} = [trapezium, trapezium left angle=70, trapezium right angle=110, minimum width=3cm, minimum height=1cm, text centered, draw=black, fill=blue!30]
\tikzstyle{process} = [rectangle, minimum width=3cm, minimum height=1cm, text centered, draw=black, fill=orange!30]
\tikzstyle{decision} = [diamond, minimum width=3cm, minimum height=1cm, text centered, draw=black, fill=green!30]
\tikzstyle{arrow} = [thick,->,>=stealth]
\ifCLASSINFOpdf
\else
\fi
\begin{document}
%
\title{LDPC Lattice Codes for Full-Duplex Relay Channels}


\author{ Hassan Khodaiemehr, Dariush Kiani and Mohammad-Reza~Sadeghi
\thanks{Hassan Khodaiemehr, Dariush Kiani  and Mohammad-Reza~Sadeghi are with the Department
of Mathematics and Computer Science, Amirkabir University of Technology (Tehran Polytechnic), Tehran, Iran. (Emails: \{h.khodaiemehr, dkiani, msadeghi\}@aut.ac.ir).
\newline Part of this work was presented in \cite{IWCIT2015} at IWCIT 2015, Iran. }
}
\maketitle

\begin{abstract}
Low density parity check (LDPC) lattices are obtained from Construction D' and a family of nested binary LDPC codes. We consider an special case of these lattices with one binary LDPC code as  underlying code. This special case of LDPC lattices can be obtained by lifting binary LDPC codes using Construction A lattices.
The LDPC lattices were the first family of lattices which have efficient decoding  in high dimensions. We employ the encoding and decoding of the LDPC lattices in a cooperative transmission framework. We establish two efficient shaping methods  based on hypercube shaping and Voronoi shaping, to obtain  LDPC lattice codes.  Then,  we propose the implementation of block Markov encoding for one-way and two-way relay networks using  LDPC lattice codes.
This entails owning an efficient method for decomposing full-rate codebook into lower rate codebooks. We apply different  decomposition schemes for one-way and two-way relay channels which  are the altered versions of the decomposition methods of low density lattice codes (LDLCs). Due to the lower complexity of the decoding for  LDPC lattices comparing to LDLCs, the complexity of
 our schemes are significantly lower than the ones  proposed for LDLCs. The efficiency of the proposed schemes are presented using simulation results.
\end{abstract}
\begin{IEEEkeywords}
Relay networks, lattice codes, block Markov encoding.
\end{IEEEkeywords}


%
\IEEEpeerreviewmaketitle
\vspace{-0.5cm}
\section{Introduction}
\PARstart{O}{ver} the last few years, cooperative transmission has become
widely prominent with the increases in the  size of communication networks.
In wireless networks, the transmitted message from a node is heard not only by its intended receiver, but also by other neighbour nodes and those neighbour nodes can use the received signals to help transmission.  They bring a
cooperative transmission by acting as relays.

The relay channel first introduced by van der Meulen in \cite{vandermullen} and it consists of a source aiming to
communicate with a destination with the help of a relay. In this case, we call the relay channel \emph{one-way relay channel} or \emph{single relay channel}. In \cite{1056084}, Cover and El Gamal proposed the fundamental \emph{decode-forward} (DF) and \emph{compress-forward} (CF) schemes for the one-way relay channels which achieve near capacity rates. In DF scheme, the relay decodes the messages from the source and forwards them to the destination. In CF scheme, the relay compresses received signals and forwards the compression indices.

It is proved that the DF scheme is optimal for these types of channels: for physically degraded relay
channels \cite{1056084} in which  the output observed at the receiver is a degraded version of the channel output at
the relay,  for semi-deterministic channels \cite{1056502} in which the channel output at the relay is a deterministic
function of the channel input at the transmitter and the channel
input at the relay. The exact capacity of general relay channels is not known to date, although, there exist tight capacity approximations for a large class of networks \cite{5730555, 5752460}, and schemes like DF and CF achieve near-capacity rates. The  upper bound on capacity is given by the cut-set upper bound \cite{1056084} and the  lower bound is given by Chong et al. in \cite{4039653}. The scheme in \cite{4039653} is a block-Markov transmission scheme that is a combination of the DF scheme and the CF scheme.

The one-way relay channel can be extended to the \emph{two-way relay channel}, where a relay helps two users exchange their messages. Two types of two-way relay channels can be considered, that is, without
a direct link between the two users, and with a direct link between the two users. The former is a suitable model for wired communication and the latter is suitable for wireless communication. Applications of relay cooperation  can be seen in increasing the capacity \cite{1246003}, combating the fading effect \cite{1362898}, mitigating the effects of interference \cite{5752442, 6560483, 6766765} and increasing the physical layer security \cite{5352243}. However, DF scheme has been used in numerous applications, it achieves capacity only in special few cases.  All of these approaches are using random Gaussian coding which is impractical for implementation. Thus, applying DF scheme in a practical scenario is interesting. One of the research areas that has such potential is lattice theory.

An $n$ dimensional lattice in $\mathbb{R}^m$ is  the set of integer linear combinations of a given basis with $n$ linearly independent vectors in $\mathbb{R}^m$. Using lattices for communication over the real AWGN channel, has been investigated by Poltyrev \cite{312163}. In such a communication system, instead of the coding rate and capacity,  normalized logarithmic density (NLD) and generalized capacity $C_{\infty}$ have been introduced, respectively.
Using Construction D of lattices \cite{3783851}, the existence of sphere-bound-achieving and capacity-achieving lattices has been proved by Forney et al. \cite{841165}. A capacity-achieving lattice code can be obtained from a capacity-achieving lattice together with a proper shaping region \cite{651040, 1337105}.
Lattice codes are the Euclidean-space analog of linear codes. Researchers have also studied practical lattice codes.
The search for practical implementable  capacity-achieving lattices and lattice codes started by proposing low density parity check  (LDPC) lattices~\cite{1705007}. In this class of lattices, a set of nested LDPC codes and Construction D' of lattices  \cite{3783851} are used to generate lattices with sparse parity check matrices. Another class of lattices, called low density lattice codes (LDLC), introduced and investigated in \cite{4475389}.
Turbo lattices employed Construction D along with turbo codes to achieve capacity gains \cite{5706882}.  Low density Construction A (LDA) lattices \cite{6404707} are another class of lattices with near-capacity error performance and low-complexity, low-storage decoder. An LDA lattice can be obtained from Construction A \cite{3783851,6555759} with a non-binary LDPC code as its underlying code.

The use of lattice codes in relay networks has received significant attentions in recent years  \cite{6034734}, \cite{6506106}, \cite{DBLP:conf/icc/FerdinandNA13} \cite{DBLP:conf/icc/NoklebyA11}, \cite{6994267}, \cite{6582523}. It was shown in \cite{DBLP:conf/icc/NoklebyA11} and \cite{6994267} that lattice codes can achieve the DF rates for the relay channel. All of these achievable schemes rely on asymptotic code lengths, which is a drawback in practical implementation.
Recently,  Aazhang et al. proposed a  practical scheme based on LDLCs, for the real-valued, full-duplex one-way and two-way relay channels \cite{6994267}.
In this work, we propose another class of practical, efficient lattice codes, based on  LDPC lattices,  for the real-valued, full-duplex one-way and two-way relay channels.

The rest of this paper is organized as follows. Section~\ref{system_model} introduces the system models of the one-way and two-way relay channels. Section~\ref{lattice} presents the preliminaries on lattices and lattice codes. In Section~\ref{LDPC Lattices}, we introduce  LDPC lattices. The encoding and decoding of these lattices are also presented in this section.
In Section~\ref{shaping_sec}, we consider the application of the  LDPC lattices in the power constrained AWGN channels by presenting two efficient shaping methods, based on hypercube shaping and nested lattice shaping.
In Section~\ref{one_way_channel}, we adapt our shaping algorithms that enable us to do the decomposition of the LDPC lattices into lower-rate components without loss of shaping efficiency. Then, we present a practical block Markov scheme for the real-valued, full-duplex one-way relay channels, based on  LDPC lattices. In Section~\ref{two_way_channel_sec}, we present another decomposition method based on doubly nested  LDPC lattices,  for the two-way relay channels. Finally, in Section~\ref{Numerical Results}, we examine the practical performance of our proposed schemes. Section~\ref{conclusions} contains the concluding remarks.
\section{System Model}\label{system_model}
\subsection{One-Way Relay Channel }
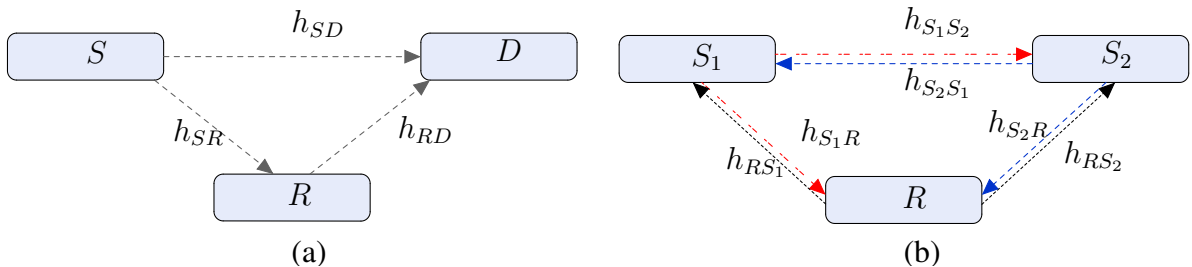
\begin{figure}[h]
\begin{center}
\definecolor{zzttqq}{rgb}{0.,0.2,0.8}
\definecolor{qqttcc}{rgb}{0.,0.2,0.8}
\definecolor{ffqqqq}{rgb}{1.,0.,0.}
\definecolor{wqwqwq}{rgb}{0.3764705882352941,0.3764705882352941,0.3764705882352941}
\begin{tikzpicture}[line cap=round,line join=round,>=triangle 45,x=0.12213740458015267cm,y=0.125cm]
\clip(5.,-57.) rectangle (136.,-25.);
\fill[line width=0.pt,color=zzttqq,fill=zzttqq,fill opacity=0.1] (67.21748791714167,-32.21948106496881) -- (52.21748791714167,-32.21948106496881) -- (51.490059965882836,-32.48807198275779) -- (51.21748791714167,-33.21948106496881) -- (51.21748791714167,-36.219481064968804) -- (51.48476550176804,-36.95578284369824) -- (52.21748791714167,-37.219481064968804) -- (67.21748791714167,-37.219481064968804) -- (67.95243514205762,-36.95561431076526) -- (68.21748791714167,-36.219481064968804) -- (68.21748791714167,-33.21948106496881) -- (67.9514822812876,-32.484145066191346) -- cycle;
\fill[line width=0.pt,color=zzttqq,fill=zzttqq,fill opacity=0.1] (22.217487917141618,-32.21948106496881) -- (7.21748791714162,-32.21948106496881) -- (6.490059965882777,-32.48807198275779) -- (6.21748791714162,-33.21948106496881) -- (6.21748791714162,-36.219481064968804) -- (6.484765501767987,-36.95578284369824) -- (7.21748791714162,-37.219481064968804) -- (22.217487917141618,-37.219481064968804) -- (22.952435142057574,-36.95561431076526) -- (23.217487917141618,-36.219481064968804) -- (23.217487917141618,-33.21948106496881) -- (22.951482281287547,-32.484145066191346) -- cycle;
\fill[line width=0.pt,color=zzttqq,fill=zzttqq,fill opacity=0.1] (44.69811370147054,-47.22790740491434) -- (29.698113701470568,-47.22790740491434) -- (28.9706857502117,-47.49649832270332) -- (28.698113701470568,-48.22790740491434) -- (28.69811370147054,-51.22790740491434) -- (28.96539128609691,-51.96420918364377) -- (29.69811370147054,-52.22790740491434) -- (44.69811370147054,-52.227907404914355) -- (45.4330609263865,-51.9640406507108) -- (45.698113701470554,-51.227907404914355) -- (45.698113701470554,-48.22790740491434) -- (45.43210806561647,-47.492571406136875) -- cycle;
\fill[line width=0.pt,color=zzttqq,fill=zzttqq,fill opacity=0.1] (88.82519949396965,-32.46041161184824) -- (73.82519949396965,-32.46041161184824) -- (73.09777154271082,-32.72900252963721) -- (72.82519949396965,-33.46041161184824) -- (72.82519949396965,-36.46041161184824) -- (73.09247707859603,-37.19671339057766) -- (73.82519949396965,-37.46041161184822) -- (88.82519949396965,-37.46041161184822) -- (89.56014671888562,-37.19654485764469) -- (89.82519949396965,-36.46041161184824) -- (89.82519949396965,-33.46041161184824) -- (89.55919385811558,-32.725075613070764) -- cycle;
\fill[line width=0.pt,color=zzttqq,fill=zzttqq,fill opacity=0.1] (133.80047395779474,-32.47152777316628) -- (118.80047395779472,-32.47152777316628) -- (118.07304600653588,-32.74011869095526) -- (117.80047395779472,-33.47152777316628) -- (117.80047395779472,-36.47152777316628) -- (118.06775154242108,-37.207829551895706) -- (118.80047395779472,-37.47152777316628) -- (133.80047395779474,-37.47152777316628) -- (134.53542118271065,-37.207661018962746) -- (134.80047395779474,-36.47152777316628) -- (134.80047395779474,-33.47152777316628) -- (134.53446832194066,-32.73619177438882) -- cycle;
\fill[line width=0.pt,color=zzttqq,fill=zzttqq,fill opacity=0.1] (111.28109974212362,-47.479954113111795) -- (96.28109974212362,-47.479954113111795) -- (95.55367179086474,-47.74854503090078) -- (95.28109974212362,-48.479954113111795) -- (95.28109974212362,-51.479954113111795) -- (95.54837732674994,-52.21625589184122) -- (96.28109974212362,-52.47995411311181) -- (111.28109974212362,-52.47995411311181) -- (112.01604696703953,-52.21608735890826) -- (112.28109974212362,-51.479954113111795) -- (112.28109974212362,-48.479954113111795) -- (112.01509410626954,-47.74461811433434) -- cycle;
\draw (7.21748791714162,-32.21948106496881)-- (22.217487917141618,-32.21948106496881);
\draw (7.21748791714162,-37.219481064968804)-- (22.217487917141618,-37.219481064968804);
\draw (6.21748791714162,-33.21948106496881)-- (6.21748791714162,-36.219481064968804);
\draw [shift={(7.21748791714162,-33.21948106496881)}] plot[domain=1.5707963267948966:3.141592653589793,variable=\t]({1.*1.*cos(\t r)+0.*1.*sin(\t r)},{0.*1.*cos(\t r)+1.*1.*sin(\t r)});
\draw (23.217487917141618,-33.21948106496881)-- (23.217487917141618,-36.219481064968804);
\draw [shift={(7.21748791714162,-36.219481064968804)}] plot[domain=3.141592653589793:4.71238898038469,variable=\t]({1.*1.*cos(\t r)+0.*1.*sin(\t r)},{0.*1.*cos(\t r)+1.*1.*sin(\t r)});
\draw [shift={(22.217487917141618,-33.21948106496881)}] plot[domain=0.:1.5707963267948966,variable=\t]({1.*1.*cos(\t r)+0.*1.*sin(\t r)},{0.*1.*cos(\t r)+1.*1.*sin(\t r)});
\draw [shift={(22.217487917141618,-36.219481064968804)}] plot[domain=-1.5707963267948966:0.,variable=\t]({1.*1.*cos(\t r)+0.*1.*sin(\t r)},{0.*1.*cos(\t r)+1.*1.*sin(\t r)});
\draw (52.21748791714167,-32.21948106496881)-- (67.21748791714167,-32.21948106496881);
\draw (52.21748791714167,-37.219481064968804)-- (67.21748791714167,-37.219481064968804);
\draw (51.21748791714167,-33.21948106496881)-- (51.21748791714167,-36.219481064968804);
\draw [shift={(52.21748791714167,-33.21948106496881)}] plot[domain=1.5707963267948966:3.141592653589793,variable=\t]({1.*1.*cos(\t r)+0.*1.*sin(\t r)},{0.*1.*cos(\t r)+1.*1.*sin(\t r)});
\draw (68.21748791714167,-33.21948106496881)-- (68.21748791714167,-36.219481064968804);
\draw [shift={(52.21748791714167,-36.219481064968804)}] plot[domain=3.141592653589793:4.71238898038469,variable=\t]({1.*1.*cos(\t r)+0.*1.*sin(\t r)},{0.*1.*cos(\t r)+1.*1.*sin(\t r)});
\draw [shift={(67.21748791714167,-33.21948106496881)}] plot[domain=0.:1.5707963267948966,variable=\t]({1.*1.*cos(\t r)+0.*1.*sin(\t r)},{0.*1.*cos(\t r)+1.*1.*sin(\t r)});
\draw [shift={(67.21748791714167,-36.219481064968804)}] plot[domain=-1.5707963267948966:0.,variable=\t]({1.*1.*cos(\t r)+0.*1.*sin(\t r)},{0.*1.*cos(\t r)+1.*1.*sin(\t r)});
\draw (29.698113701470568,-47.22790740491434)-- (44.69811370147054,-47.22790740491434);
\draw (29.698113701470568,-52.207120204149255)-- (44.69811370147054,-52.227907404914355);
\draw (28.698113701470568,-48.22790740491434)-- (28.698113701470568,-51.207120204149255);
\draw [shift={(29.698113701470568,-48.22790740491434)}] plot[domain=1.5707963267948966:3.141592653589793,variable=\t]({1.*1.*cos(\t r)+0.*1.*sin(\t r)},{0.*1.*cos(\t r)+1.*1.*sin(\t r)});
\draw (45.698113701470554,-48.22790740491434)-- (45.698113701470554,-51.227907404914355);
\draw [shift={(29.698113701470568,-51.207120204149255)}] plot[domain=3.141592653589793:4.71238898038469,variable=\t]({1.*1.*cos(\t r)+0.*1.*sin(\t r)},{0.*1.*cos(\t r)+1.*1.*sin(\t r)});
\draw [shift={(44.69811370147054,-48.22790740491434)}] plot[domain=0.:1.5707963267948966,variable=\t]({1.*1.*cos(\t r)+0.*1.*sin(\t r)},{0.*1.*cos(\t r)+1.*1.*sin(\t r)});
\draw [shift={(44.69811370147054,-51.227907404914355)}] plot[domain=-1.5707963267948966:0.,variable=\t]({1.*1.*cos(\t r)+0.*1.*sin(\t r)},{0.*1.*cos(\t r)+1.*1.*sin(\t r)});
\draw [->,dash pattern=on 2pt off 2pt,color=wqwqwq] (23.217487917141618,-34.719481064968804) -- (51.21748791714167,-34.72874555724628);
\draw (13.580557760431486,-31.894304393840127) node[anchor=north west] {$S$};
\draw (58.151882220748504,-31.74423259431044) node[anchor=north west] {$D$};
\draw (35.49104049176577,-47.05155614633853) node[anchor=north west] {$R$};
\draw [->,dash pattern=on 2pt off 2pt,color=wqwqwq] (22.217487917141618,-37.219481064968804) -- (35.180406011112964,-47.22790740491434);
\draw [->,dash pattern=on 2pt off 2pt,color=wqwqwq] (39.13768193920707,-47.22790740491434) -- (52.21748791714167,-37.219481064968804);
\draw (35.94125589035484,-28.592724804187007) node[anchor=north west] {$h_{SD}$};
\draw (23.03508113080176,-40.14825336797292) node[anchor=north west] {$h_{SR}$};
\draw (47.49678445414073,-39.848109768913545) node[anchor=north west] {$h_{RD}$};
\draw (35.94125589035484,-53.054428127526016) node[anchor=north west] {(a)};
\draw (73.80047395779467,-32.47152777316628)-- (88.80047395779465,-32.47152777316628);
\draw (73.80047395779467,-37.47152777316628)-- (88.80047395779465,-37.47152777316628);
\draw (72.80047395779467,-33.47152777316628)-- (72.80047395779467,-36.47152777316628);
\draw [shift={(73.80047395779467,-33.47152777316628)}] plot[domain=1.5707963267948966:3.141592653589793,variable=\t]({1.*1.*cos(\t r)+0.*1.*sin(\t r)},{0.*1.*cos(\t r)+1.*1.*sin(\t r)});
\draw (89.80047395779465,-33.47152777316628)-- (89.80047395779465,-36.47152777316628);
\draw [shift={(73.80047395779467,-36.47152777316628)}] plot[domain=3.141592653589793:4.71238898038469,variable=\t]({1.*1.*cos(\t r)+0.*1.*sin(\t r)},{0.*1.*cos(\t r)+1.*1.*sin(\t r)});
\draw [shift={(88.80047395779465,-33.47152777316628)}] plot[domain=0.:1.5707963267948966,variable=\t]({1.*1.*cos(\t r)+0.*1.*sin(\t r)},{0.*1.*cos(\t r)+1.*1.*sin(\t r)});
\draw [shift={(88.80047395779465,-36.47152777316628)}] plot[domain=-1.5707963267948966:0.,variable=\t]({1.*1.*cos(\t r)+0.*1.*sin(\t r)},{0.*1.*cos(\t r)+1.*1.*sin(\t r)});
\draw (118.80047395779472,-32.47152777316628)-- (133.80047395779474,-32.47152777316628);
\draw (118.80047395779472,-37.47152777316628)-- (133.80047395779474,-37.47152777316628);
\draw (117.80047395779472,-33.47152777316628)-- (117.80047395779472,-36.47152777316628);
\draw [shift={(118.80047395779472,-33.47152777316628)}] plot[domain=1.5707963267948966:3.141592653589793,variable=\t]({1.*1.*cos(\t r)+0.*1.*sin(\t r)},{0.*1.*cos(\t r)+1.*1.*sin(\t r)});
\draw (134.80047395779474,-33.47152777316628)-- (134.80047395779474,-36.47152777316628);
\draw [shift={(118.80047395779472,-36.47152777316628)}] plot[domain=3.141592653589793:4.71238898038469,variable=\t]({1.*1.*cos(\t r)+0.*1.*sin(\t r)},{0.*1.*cos(\t r)+1.*1.*sin(\t r)});
\draw [shift={(133.80047395779474,-33.47152777316628)}] plot[domain=0.:1.5707963267948966,variable=\t]({1.*1.*cos(\t r)+0.*1.*sin(\t r)},{0.*1.*cos(\t r)+1.*1.*sin(\t r)});
\draw [shift={(133.80047395779474,-36.47152777316628)}] plot[domain=-1.5707963267948966:0.,variable=\t]({1.*1.*cos(\t r)+0.*1.*sin(\t r)},{0.*1.*cos(\t r)+1.*1.*sin(\t r)});
\draw (96.28109974212362,-47.479954113111795)-- (111.28109974212362,-47.479954113111795);
\draw (96.28109974212362,-52.47995411311181)-- (111.28109974212362,-52.47995411311181);
\draw (95.28109974212362,-48.479954113111795)-- (95.28109974212362,-51.479954113111795);
\draw [shift={(96.28109974212362,-48.479954113111795)}] plot[domain=1.5707963267948966:3.141592653589793,variable=\t]({1.*1.*cos(\t r)+0.*1.*sin(\t r)},{0.*1.*cos(\t r)+1.*1.*sin(\t r)});
\draw (112.28109974212362,-48.479954113111795)-- (112.28109974212362,-51.479954113111795);
\draw [shift={(96.28109974212362,-51.479954113111795)}] plot[domain=3.141592653589793:4.71238898038469,variable=\t]({1.*1.*cos(\t r)+0.*1.*sin(\t r)},{0.*1.*cos(\t r)+1.*1.*sin(\t r)});
\draw [shift={(111.28109974212362,-48.479954113111795)}] plot[domain=0.:1.5707963267948966,variable=\t]({1.*1.*cos(\t r)+0.*1.*sin(\t r)},{0.*1.*cos(\t r)+1.*1.*sin(\t r)});
\draw [shift={(111.28109974212362,-51.479954113111795)}] plot[domain=-1.5707963267948966:0.,variable=\t]({1.*1.*cos(\t r)+0.*1.*sin(\t r)},{0.*1.*cos(\t r)+1.*1.*sin(\t r)});
\draw [->,dash pattern=on 1pt off 1pt on 2pt off 4pt,color=ffqqqq] (81.80047395779465,-37.47152777316628) -- (95.28109974212362,-49.4315279994261);
\draw [->,dash pattern=on 1pt off 1pt] (95.28109974212362,-50.40898585229985) -- (80.80047395779465,-37.47152777316628);
\draw [->,dash pattern=on 1pt off 1pt] (112.28109974212362,-50.449155353102896) -- (126.80047395779474,-37.47152777316628);
\draw [->,dash pattern=on 2pt off 2pt,color=qqttcc] (125.80047395779471,-37.47152777316628) -- (112.28109974212362,-49.444917833027134);
\draw [->,dash pattern=on 1pt off 1pt on 2pt off 4pt,color=ffqqqq] (89.80047395779465,-34.47152777316628) -- (117.80047395779472,-34.47152777316628);
\draw [->,dash pattern=on 2pt off 2pt,color=qqttcc] (117.80047395779472,-35.47152777316628) -- (89.80047395779465,-35.47152777316628);
\draw (79.46207775396404,-32.1944479928995) node[anchor=north west] {$S_1$};
\draw (124.03340221428105,-31.894304393840127) node[anchor=north west] {$S_2$};
\draw (102.42306308200614,-47.3516997453979) node[anchor=north west] {$R$};
\draw (102.57313488153582,-28.292581205127632) node[anchor=north west] {$h_{S_{1}S_2}$};
\draw (102.57313488153582,-34.74566858490418) node[anchor=north west] {$h_{S_{2}S_1}$};
\draw (91.3177499168093,-40.14825336797292) node[anchor=north west] {$h_{S_1R}$};
\draw (83.21387274220622,-43.14968935856666) node[anchor=north west] {$h_{RS_1}$};
\draw (111.72751465284674,-39.247822570794796) node[anchor=north west] {$h_{S_2R}$};
\draw (119.9814636269795,-42.54940216044791) node[anchor=north west] {$h_{RS_2}$};
\draw (102.57313488153582,-53.054428127526016) node[anchor=north west] {(b)};
\end{tikzpicture}
\caption{(a) Full-duplex Gaussian one-way relay channel.  (b) Full-duplex Gaussian two-way relay channel.}
\label{channel_model}
\end{center}
\end{figure}

The relay channel that we have considered, is a three-terminal relay channel. First, we present the one-way relay channel, depicted in  \figurename{\ref{channel_model}}~(a).  The source transmits a message, which is mapped to a codeword, to both relay and destination and in the next time slot relay aids the destination by sending the part of the information of the previous time slot. We assume a full-duplex relay which can simultaneously transmit and receive the massages. For simplicity we suppose real-valued channels.
Let $\mathbf{x}_S$ and $\mathbf{x}_R$ denote the signals transmitted by the source and the relay. Let $\mathbf{y}_R$ and $\mathbf{y}_D$ denote the signals received at the relay and the destination. The received signals are
\begin{eqnarray}
  \mathbf{y}_R &=& h_{SR}\mathbf{x}_S+\mathbf{z}_R \\
  \mathbf{y}_D &=& h_{SD}\mathbf{x}_S+h_{RD}\mathbf{x}_{R}+\mathbf{z}_{D},
\end{eqnarray}
where $\mathbf{z}_R\sim \mathcal{N}(0,N_R)$, $\mathbf{z}_D\sim \mathcal{N}(0,N_D)$. Moreover, $h_{SR}= d_{SR}^{-\alpha_1}$, $h_{SD}=1$, and $h_{RD}=d_{RD}^{-\alpha_2}$ are the channel gains between source, relay, and destination, which follow the usual path-loss model. In particular, we take the distance between source and destination to be unity, $d_{SR}$ and $d_{RD}$ to be the distance between source to relay and relay to destination, and $\alpha_1$ and $\alpha_2$ to be the corresponding path-loss exponents. We constrain the source and relay transmissions to have average power no greater than $P_S$ and $P_R$.

In general, the capacity of this channel is unknown however, the DF scheme which is proposed in \cite{1056084}, achieves the following inner bound:
\begin{eqnarray}\label{rate_bound}
  R\leq \frac{1}{2} \min \left\{ \log_2 \left(1+\frac{h_{SR}^2P_SE\left\{\mathbf{x}_S^2\right\}}{N_R} \right),  \log_2 \left(1+\frac{h_{SD}^2P_S E\left\{\mathbf{x}_S^2\right\}+h_{RD}^2P_RE\left\{\mathbf{x}_R^2\right\} }{N_D} \right)  \right\}.\nonumber
\end{eqnarray}
This rate can be achieved via \emph{block Markov encoding}.  After decoding the message, the relay re-encodes the message and transmits the corresponding codeword in the next block.
The lattice-coding version of block Markov encoding is proposed in \cite{6506106}. It is proved theoretically that it can achieve the decode-and-forward rates. The results of \cite{6506106} suggest that structured lattice codes may be used
to outperform, random Gaussian codes in general Gaussian networks.  The authors of \cite{6994267} have applied this scheme and designed a family of practically implementable LDLC lattice codes for relay channels. In this paper we present another family of lattice codes which are amenable to practical implementation in block Markov schemes.
\subsection{Two-Way Relay Channel}
\makeatletter 
    \def\tagform@#1{\maketag@@@{\normalsize(#1)\@@italiccorr}}
\makeatother
Next, we present the full-duplex Gaussian two-way relay channel, as depicted in  \figurename{\ref{channel_model}}~(b). The two sources $S_1$ and $S_2$ exchange their messages, which are mapped to codewords and transmitted over the wireless medium. The relay node $R$ overhears the noisy superposition of signals transmitted from sources and makes its own transmissions to facilitate communications. This channel can be modeled as follows
\begin{eqnarray}\label{two_way_channel}
  \mathbf{y}_R &=& h_{S_1R}\mathbf{x}_{S_1}+h_{S_2R}\mathbf{x}_{S_2}+\mathbf{z}_R, \\
  \mathbf{y}_{S_1} &=& h_{S_2S_1}\mathbf{x}_{S_2}+h_{RS_1}\mathbf{x}_R, \\
  \mathbf{y}_{S_2} &=& h_{S_1S_2}\mathbf{x}_{S_1}+h_{RS_2}\mathbf{x}_R,
\end{eqnarray}
where the noise components are $\mathbf{z}_R\sim \mathcal{N}(0,N_R)$, $\mathbf{z}_{S_1}\sim \mathcal{N}(0,N_{S_1})$ and $\mathbf{z}_{S_2}\sim \mathcal{N}(0,N_{S_2})$. Similar to the one-way relay
channel model, $h_{S_1S_2}$, $h_{S_2S_1}$, $h_{S_2R}$, $h_{S_1R}$, $h_{RS_1}$ and $h_{RS_2}$ are the channel gains which follow the usual path-loss model.  We constrain the sources $S_1$, $S_2$ and relay $R$ transmissions to have average powers no greater than $P_{S_1}$, $P_{S_2}$ and $P_R$, respectively. The capacity of this channel is unknown, but a DF scheme was presented in~\cite{6506106} that achieves rate pairs $(R_1,R_2)$ satisfying (\ref{rate_bound2}),
\begin{align}\label{rate_bound2}
 R_i & \leq \min \left\{ \frac{1}{2}\log_2\left(1+\frac{h_{RS_j}^2P_R E\left\{\mathbf{x}_{R}^2\right\}+h_{S_iS_j}^2P_{S_i}E\left\{\mathbf{x}_{S_i}^2\right\}}{N_{S_j}}\right),\right. \nonumber\\
  &\quad \left.\left[ \frac{1}{2}\log_2\left(\frac{h_{S_iR}^2 P_{S_i}E\left\{\mathbf{x}_{S_i}^2\right\} }{h_{S_iR}^2 P_{S_i}E\left\{\mathbf{x}_{S_i}^2\right\}+h_{S_jR}^2P_{S_j}E\left\{\mathbf{x}_{S_j}^2\right\}}+\frac{h_{S_iR}^2 P_{S_i}E\left\{\mathbf{x}_{S_i}^2\right\}}{N_R}\right)\right]^{+} \right\},
\end{align}
in which $[x]^{+}\triangleq \max \left\{x,0\right\}$,
 $i,j\in\left\{1,2\right\}$, and $i\neq j$.
\section{Preliminaries on Lattices}~\label{lattice}
A discrete, additive subgroup $\Lambda$ of the $m$-dimensional real space $\mathbb{R}^m$ is  a lattice.
Every lattice $\Lambda$ has a basis $\mathcal{B}=\{\mathbf{b}_1,\ldots,\mathbf{b}_n\}\subset\mathbb{R}^m$ where every $ \mathbf{x}\in\Lambda$ can be represented as an integer linear combination of vectors in $\mathcal{B}$. The rank of the lattice is $n$ and its dimension is $m$. If $n = m$, the lattice is called a full-rank lattice. In this paper, we consider full-rank lattices. The matrix $\mathbf{M}$ with $\mathbf{b}_1,\ldots, \mathbf{b}_n$ as rows, is a generator matrix
for the lattice. The matrix $\mathbf{G} = \mathbf{M}\mathbf{M}^t$ is a Gram matrix for
the lattice. The determinant of the lattice $\det(\Lambda)$ is defined to be the determinant of the matrix $\mathbf{G}$ and
the volume of the lattice is defined as $\textrm{vol}(\Lambda) =\sqrt{\det(\mathbf{G})}$.
A Voronoi cell $\mathcal{V}(\mathbf{x})$ is the set of those points of $\mathbb{R}^n$ that are at least as close to $\mathbf{x}$ as to any other point in $\Lambda$. We call the Voronoi region associated with the origin,
the fundamental Voronoi region of $\Lambda$, denoted by $\mathcal{V}$ or $\mathcal{V}(\Lambda)$.

We say that a lattice $\Lambda_s$ is nested in $\Lambda_c$ if $\Lambda_s \subset \Lambda_c$. Using
nested lattices in $\mathbb{R}^n$, define the codebook $\mathcal{C}=\mathcal{V}_s\cap \Lambda_c$ which has the rate
\begin{equation}\label{rate_def_nested}
  R=\frac{1}{n}\log_2(|\mathcal{C}|)=\frac{1}{n} \log_2\left(\frac{\textrm{vol}(\mathcal{V}_s)}{\textrm{vol}(\mathcal{V}_c)}\right).
\end{equation}
Suppose that the points of a lattice $\Lambda$ are sent over an unconstrained AWGN channel with noise variance $\sigma^2$. The volume-to-noise ratio (VNR) of lattice $\Lambda$ is defined as
\begin{equation}~\label{VNR}
\textrm {VNR}=\frac{\textrm{vol}(\Lambda)^{\frac{2}{n}}}{2\pi e\sigma^2}.
\end{equation}
For a large  $n$, VNR is the ratio of the normalized volume of $\Lambda$ to the normalized volume of a noise sphere of squared radius $n\sigma^2$ which is defined as generalized signal-to-noise ratio (SNR) in~\cite{1705007} and $\alpha^2$ in \cite{841165}.

Let $\mathbf{x}\in\Lambda$ be the transmitted vector on the unconstrained AWGN channel, then the received vector $\mathbf{r}$ can be written as $\mathbf{r}=\mathbf{x}+\mathbf{e}$,
where $\mathbf{e}=(e_1,\ldots,e_n)$ and its components are independently and identically distributed (i.i.d.) Gaussian random variables with zero mean and variance $\sigma^2$. The probability of correct decoding, under maximum likelihood decoding, is given by
\begin{equation}~\label{pe}
P_c=\frac{1}{(\sigma\sqrt{2\pi})^n}\int_{\mathcal{V}(\mathbf{x})}e^{\frac{-\|\mathbf{t}\|^2}{2\sigma^2}}d{\mathbf{t}},
\end{equation}
where $\| \mathbf{x}\|$ is the Euclidean norm of $\mathbf{x}$.
A lattice quantizer is a map $\mathcal{Q}_{\Lambda}:\mathbb{R}^n\rightarrow \Lambda$ for some lattice $\Lambda \subset \mathbb{R}^n$. If we use the nearest-neighbor quantizer $\mathcal{Q}_{\Lambda}^{(NN)}$, then the
quantization error $\mathbf{x}_e\triangleq \mathbf{x}-\mathcal{Q}_{\Lambda}^{(NN)}(\mathbf{x})\in\mathcal{V}(\Lambda)$. Let $\mathbf{x}_e$ be uniformly distributed over the Voronoi region $\mathcal{V}(\Lambda)$. Then, the \emph{second moment per dimension} of $\Lambda$ is
\begin{equation}\label{second_moment}
  \sigma^2(\Lambda)=E\left[\|\mathbf{x}_e\|^2\right]=\frac{1}{n}\frac{1}{\textrm{vol}(\Lambda)}\int_{\mathcal{V}(\Lambda)}\|\mathbf{x}_e\|^2d\mathbf{x}_e.
\end{equation}

A \emph{lattice constellation} $C(\Lambda,\mathcal{R})=(\Lambda +\mathbf{t})\cap \mathcal{R}$ is the finite set of points in a lattice translate $\Lambda + \mathbf{t}$ that lies within a compact bounding region $\mathcal{R}$ of $n$-dimensional real space $\mathbb{R}^n$. The key geometric properties
of the region $\mathcal{R}$ are its \emph{volume} $\textrm{vol}(\mathcal{R})$ and the \emph{average energy} $P(\mathcal{R})$ per dimension of a uniform probability density function over $\mathcal{R}$ (see, e.g. \cite{720542} and \cite{3783851}):
\begin{equation}\label{eq1}
    P(\mathcal{R})=\int_\mathcal{R} \frac{(\|\mathbf{x} \|^2/n)d\mathbf{x}}{\textrm{vol}(\mathcal{R})}.
\end{equation}
The \emph{normalized second moment} of $\mathcal{R}$ is defined as
\begin{equation}\label{eq2}
    G(\mathcal{R})=\frac{P(\mathcal{R})}{\textrm{vol}(\mathcal{R})^{2/n}}.
\end{equation}
The normalized second moment of any $n$-cube centered at the origin is $1/12$. The \emph{shaping gain} $\gamma_s(\mathcal{R})$ of $\mathcal{R}$, measures the decrease in average energy of $\mathcal{R}$ relative to a baseline region, namely, an interval $[-d_0/2, d_0/2]$ or an $n$-cube $[-d_0/2, d_0/2]^n$, where $d_0$ is related to the $\textrm{vol}(\mathcal{R})$ \cite{3783851}. The definition of shaping gain is
\begin{equation}\label{eq3}
    \gamma_s(\mathcal{R})=\frac{\textrm{vol}(\mathcal{R})^{2/n}}{12 P(\mathcal{R})}=\frac{1}{12G(\mathcal{R})}.
\end{equation}
The optimum $n$-dimensional shaping region is an $n$-sphere~\cite{720542}. The key geometrical parameters of an $n$-sphere $(=\otimes)$
of radius $r$ for an even $n$ are \cite{3783851}:
\begin{eqnarray}
  \textrm{vol}(\otimes) &=& \frac{(\pi r^2)^{n/2}}{(n/2)!}, \label{sphereV}\\
  P(\otimes) &=& \frac{r^2}{n+2}\label{sphereP},\\
  G(\otimes) &=& \frac{P(\otimes)}{\textrm{vol}(\otimes)^{2/n}}=\frac{((n/2)!)^{2/n}}{\pi(n+2)}. \label{sphereG}
\end{eqnarray}
The shaping gain of an $n$-sphere is a function of the dimension $n$.
When $n$ approaches infinity  the shaping gain approaches the \emph{ultimate shaping gain} $\pi e /6$ ($1.53$dB). The \emph{shaping loss} $\lambda_{s} (\mathcal{R})$ of a shaping region $\mathcal{R}$ with respect to an $n$-dimensional sphere, where $n$ is even, based on (\ref{sphereV})-(\ref{sphereG}),  is~\cite{5205636}
\begin{equation}\label{eq6}
    \lambda_{s}(\mathcal{R})=\frac{G(\mathcal{R})}{G(\otimes)}=\frac{\pi (n+2)G(\mathcal{R})}{\Gamma(\frac{n}{2}+1)^{2/n}}.
\end{equation}
The shaping loss is greater than or equal to $1$. If we form the intersection $\Lambda\cap \mathcal{R}$ of a lattice with a shaping region $\mathcal{R}\subset \mathbb{R}^n$, we would expect to obtain a code with about $\textrm{vol}(\mathcal{R})/\textrm{vol}(\Lambda)$  codewords~\cite{641543}. In fact, by using Minkowski-Hlawka~Theorem, it is proved  that the value $\textrm{vol}(\mathcal{R})/\textrm{vol}(\Lambda)$ is correct in the average over a suitable set of lattices based on codes~\cite{641543}. The rate of the code $\mathcal{C}=\Lambda\cap \mathcal{R}$ is approximately
\begin{equation}\label{rate_loeliger}
  R\approx\frac{1}{n}\log_2\left(\frac{\textrm{vol}(\mathcal{R})}{\textrm{vol}(\Lambda)}\right).
\end{equation}
\section{LDPC Lattices}~\label{LDPC Lattices}
There exist many ways to construct lattices based on codes \cite{3783851}. Here we mention two of them. The first one is Construction A and the other one is Construction D'.
Assume that $\mathcal{C}$ is a linear code over $\mathbb{F}_p$ where $p$ is a prime number, i.e. $\mathcal{C}\subseteq\mathbb{F}_{p}^n$. A lattice $\Lambda$ based on Construction A~\cite{3783851} can be derived from $\mathcal{C}$ as follows
\begin{equation}\label{constA}
\Lambda=p\mathbb{Z}^n+\epsilon\left(\mathcal{C}\right),
\end{equation}
where $\epsilon\colon\mathbb{F}_{p}^n\rightarrow\mathbb{R}^n$ is the embedding function. In this work, we are particularly interested in lattices with $p=2$.

Construction D' converts a set of parity checks defined by a family of nested codes $\mathcal{C}_0\supseteq \mathcal{C}_1\supseteq \cdots \supseteq \mathcal{C}_a$, into congruences for a lattice~\cite{3783851}. The number $a+1$ is called the level of the construction.
An LDPC lattice $\Lambda\subset \mathbb{Z}^n$ can be constructed from Construction D' and a number of nested binary LDPC codes.
More detail about the structure and decoding of these lattices can be found in \cite{1705007}. If we consider one code as underlying code of Construction D', which means $a=0$, Construction A is obtained~\cite[Proposition 1]{6555759}. In this case, Construction A LDPC lattices or 1-level LDPC lattices \cite{6555759} will be obtained. In this paper, we refer to them as LDPC lattices without mentioning the level of the construction.
\begin{Definition}
An  LDPC lattice $\Lambda$ is a lattice constructed based on Construction A or D' along with one binary  LDPC code $\mathcal{C}$ as its underlying code. Equivalently, $\mathbf{x}\in\mathbb{Z}^n$ is in $\Lambda$ if and only if $\mathbf{H}\mathbf{x}^t=\mathbf{0} \pmod{2}$, where $\mathbf{H}$ is the parity check matrix of $\mathcal{C}$.
\end{Definition}

The Generator matrix of Construction A lattice $\Lambda$ using the underlying code $\mathcal{C}\subset \mathbb{F}_2^n$ is of the form
\begin{eqnarray}\label{eq12}
  \mathbf{G}_{\Lambda} &=& \left[
                    \begin{array}{cc}
                      \textbf{I}_{k}& \mathbf{P}_{k\times (n-k)} \\
                      \textbf{0}_{(n-k)\times k} & 2\textbf{I}_{n-k} \\
                    \end{array}
                  \right],
\end{eqnarray}
where $\mathbf{G}_{\mathcal{C}}=\left[
                           \begin{array}{cc}
                              \textbf{I}_k & \mathbf{P}\\
                           \end{array}
                         \right]$
is the generator matrix of $\mathcal{C}$ in systematic form, $k$ is the rank of $\mathcal{C}$, $\textbf{I}_k$ and $\textbf{0}_k$, are identity and all zero square matrices of size $k$, respectively.
\subsection{Encoding and Decoding of  LDPC Lattices}
The practical encoding and decoding of LDPC lattices, both with linear complexity in the dimension of the lattice, has been addressed in \cite{QC-LDPC}.
In this paper, we consider a translated and scaled version of the lattice $\Lambda$, generated by (\ref{eq12}), as suggested in \cite{1057135,QC-LDPC} and \cite[\S 20.5]{3783851}.
In the sequel, we present the  decoding of these scaled and translated versions of  LDPC lattices,
which is proposed in \cite{QC-LDPC} and it is obtained  by combining the suggested decoding method in \cite[\S 20.5]{3783851}, and the decoding of binary LDPC codes.
Construction and decoding of these new lattices can be done using the following steps. First, convert the codewords of $[n, k]$~binary code $\mathcal{C}$ into $\pm 1$ notation (convert $0$ to $-1$ and $1$ to $1$) \cite[\S 20.5]{3783851} which produces a set $\Lambda(\mathcal{C})$ consisting of the vectors of the form
\begin{equation}\label{newlattice}
  \mathbf{c}+4\mathbf{z}, \quad \mathbf{c}\in \mathcal{C},\,\, \mathbf{z}\in \mathbb{Z}^n.
\end{equation}
The set of points in (\ref{newlattice}) strictly speaking, is  not a lattice, but the translate of a lattice by the vector $(-1, -1,\ldots , -1)$. The regular addition of vectors of the form (\ref{newlattice}) will not be of the same form. However, we can show that $\Lambda(\mathcal{C})$ is closed under following addition. For any $\boldsymbol{\lambda}_1,\boldsymbol{\lambda}_2\in\Lambda(\mathcal{C})$, we have \cite{IWCIT2015,QC-LDPC}
\begin{equation}\label{sum}
  \boldsymbol{\lambda}_1\oplus\boldsymbol{\lambda}_2\triangleq\boldsymbol{\lambda}_1+\boldsymbol{\lambda}_2+(1,\ldots ,1)\in\Lambda(\mathcal{C}).
\end{equation}
The encoding of an integer row vector $\mathbf{b}\in \mathbb{Z}^n$ can be done as follows
\begin{equation}\label{encoding}
  \mathcal{E}(\mathbf{b})=2\mathbf{bG}_{\Lambda}-(1,\ldots ,1),
\end{equation}
where $\mathcal{E}$ is encoding function and $\mathbf{G}_{\Lambda}$ is defined as (\ref{eq12}).
Let $\mathbf{x}=\mathbf{c}+4\mathbf{z}$ be the transmitted lattice vector and  $\mathbf{y}$ be the received  vector from AWGN channel
\begin{equation}\label{AWGN_output}
\mathbf{y}=\mathbf{c}+4\mathbf{z}+\mathbf{n},
\end{equation}
where $\mathbf{c}\in\mathcal{C}$  and $\mathcal{C}$ is a binary LDPC code in $\pm 1$ notation, $\mathbf{z}\in \mathbb{Z}^n$ and $\mathbf{n}\sim \mathcal{N}(0,\sigma^2)$. First, we decode $\mathbf{c}$ and next we find $\mathbf{z}$. The proposed algorithm in \cite{QC-LDPC} is similar to the sum-product algorithm (SPA) for LDPC codes in message passing structure~\cite{sara}.
The inputs are the log likelihood ratios (LLR) for the a priori message probabilities from each channel. The estimation of  the LLR vector $\boldsymbol{\gamma}=(\gamma_1,\ldots,\gamma_n)$ for LDPC lattices is proposed in \cite{QC-LDPC} as follows
\begin{eqnarray}\label{LLR}
\gamma_i =\log \left(\frac{\textrm{Pr}\left\{c_i=-1|y_i\right\}}{\textrm{Pr}\left\{c_i=+1|y_i\right\}}\right)
&\triangleq&2\left(\frac{ (\frac{y_i-1}{4}-\lfloor(\frac{y_i-1}{4})\rceil  )^2-  (\frac{y_i+1}{4}-\lfloor(\frac{y_i+1}{4})\rceil )^2}{\sigma^2}\right),
\end{eqnarray}
where $\lfloor x\rceil$ is the nearest  integer to $x$. Input the LLR vector $\boldsymbol{\gamma}=(\gamma_1,\ldots, \gamma_n)$ to SPA decoder of LDPC codes and consider $\hat{\mathbf{c}}$ as the output of this decoder. Convert $\hat{\mathbf{c}}$ to $\pm 1$ notation and call the obtained vector $\hat{\mathbf{c}}'$. Estimate $\hat{\mathbf{z}}$ as follows
\begin{equation}\label{z_hat}
\hat{\mathbf{z}}=\left\lfloor\frac{\mathbf{y}}{4}-\frac{\hat{\mathbf{c}}'}{4}\right\rceil.
\end{equation}
Then, $\hat{\mathbf{x}}=\hat{\mathbf{c}}'+4\hat{\mathbf{z}}$ is the final decoded lattice vector.

The complexity of this decoding algorithm is significantly lower than other  lattices with practical decoding algorithm like LDA lattices \cite{6404707} and LDLCs \cite{4475389}.
The decoding algorithm of LDLCs first proposed in~\cite{4475389} which has complexity $O(n\cdot d\cdot t\cdot \frac{1}{\Delta}\cdot \log_2(\frac{1}{\Delta}))$, where $\Delta$ is the resolution and its typical value is $1/256$, $n$ is the dimension of lattice, $t$ is the number of iterations and $d$ is the average code degree. Then, in~\cite{5205636}, a new algorithm proposed with lower complexity $O(n\cdot d\cdot t\cdot K\cdot M^3)$, where $K$ is the number of replications, $n$, $t$  and $d$ are similar to above. Proposed typical value for $K$ is $3$ and for $M$ it is $2$ or $6$.  The computational complexity of the proposed decoding algorithm in \cite{QC-LDPC} is only $O(n\cdot d\cdot t)$, which is significantly lower than the complexity of LDLCs' decoding.

The main purpose of this section, is constructing  practically implementable lattice constellations. As we mentioned in previous section, definition of lattice constellations entails finding the intersection of shaping region and a translated infinite lattice. Thus, translating by $(-1,\ldots,-1)$ in the proposed structure has no inconsistency with the definition of lattice codes.
In the next section, we present efficient shaping methods to obtain a family of lattice codes based on  LDPC lattices.
\section{Shaping Methods for LDPC lattices}\label{shaping_sec}
\subsection{Hypercube Shaping Method}
In practical channels, there exists a power constraint which is needed to be  fulfilled. This entails selecting a group of lattice points with a bounded norm. In theoretical approaches, the coding lattice is intersected with a spherical shaping region to produce an efficient, power-constrained lattice code. However, spherical shaping has high computational complexity both for encoding and decoding.  In \cite{5351439}, several efficient and practical shaping algorithms proposed for LDLC lattices. Here, we present an efficient and practical shaping algorithm  for  LDPC lattices based on hypercube shaping method.

In order to perform shaping, restrict  the integer vector $\mathbf{b}$ to the following finite constellation
\begin{eqnarray}\label{constelation}
  b_{i} \in \mathcal{L}_{i}= \left\{\frac{-L_{i}}{2},\ldots , \frac{L_{i}}{2}-1 \right\},\quad\quad  i=1,\ldots, n.
\end{eqnarray}
Shape the lattice codeword $\mathbf{x}=\mathbf{bG}_{\Lambda}$ by translating each $b_i$ by an integer multiple of $L_i$, $i=1,\ldots , n$. This is equivalent to  transmitting a lattice point $\mathbf{x}'$ as follows
\begin{equation}\label{shaping}
  \mathbf{x}'=(\mathbf{b}-\mathbf{sL})\mathbf{G}_{\Lambda}=\mathbf{x}-\mathbf{sLG}_{\Lambda},
\end{equation}
where $\mathbf{L}=\textrm{diag}(L_1,\ldots ,L_n)$ is a diagonal matrix  and the new integer vector is $\mathbf{b}'=\mathbf{b}-\mathbf{sL}$. The choice of integer
vector $\mathbf{s}$, depends on the employed shaping method. In hypercube shaping we choose $\mathbf{s}$ such that the new lattice codeword components are constrained to lie in the hypercube, i.e.  $|x_i'|\leq L_i$, for $i=1,\ldots ,n$. The authors of \cite{5351439} used a triangular structure for $\mathbf{H}$, the inverse of their generator matrix $\mathbf{G}$,  with unit diagonal elements. In this case, hypercube shaping is straightforward, and $s_i$ and $x_i$ will be found recursively. Instead, we use the generator matrix of the lattice for shaping. In our case, the generator matrix beside the triangular structure, has a simpler form which helps us to obtain the components of $\mathbf{s}$ directly. Since we have used a translated and scaled lattice, we  shape the lattice vectors inside the hypercube around the origin of coordinate. Then, we scale them by factor $2$  and translate them by the all $-1$ vector. To this end, we need to solve the following system of linear equations
\begin{equation}\label{shaping_system}
     (x_1',\ldots ,x_n') =(b_1', \ldots , b_n')
\left[
                    \begin{array}{cc}
                      \mathbf {I}_{k}& \mathbf{P}_{k\times (n-k)} \\
                      \mathbf{0}_{(n-k)\times k} & 2\mathbf{I}_{n-k} \\
                    \end{array}
                  \right],
\end{equation}
by choosing an integer vector $\mathbf{s}=(s_1,\ldots s_n)$ such that $|x_i'|\leq L_i$ for $i=1,\ldots ,n$.
From (\ref{shaping_system}), we have the following equations
\begin{equation}\label{shaping_system2}
  x_i'=\left\{\begin{array}{l}
                b_i-L_is_i, \quad\quad\quad\quad\quad\quad\quad\quad\,\,\, i=1,\ldots ,k\\
                2(b_i-L_is_i)+\sum_{j=1}^{k}P_{j,i}b_j', \quad i=k+1,\ldots , n.
              \end{array}
   \right.
\end{equation}
By choosing $s_1=s_{2}=\cdots =s_{k}=0$, we have $b_i'=b_i$ and $|x_i'|=|b_i|\leq L_i$, for $i=1,\ldots ,k$. For $i=k+1,\ldots ,n$, we have the following inequalities
\vspace{-0.3cm}
\begin{eqnarray}\label{ineq}
-L_i \leq 2b_i-2L_is_i+\sum_{j=1}^{k}P_{j,i}b_j  \leq L_i,
\end{eqnarray}
\vspace{-0.3cm}
or equivalently
 \begin{eqnarray*}
 \frac{b_i}{L_i}-\frac{1}{2}+\frac{1}{2L_i}\sum_{j=1}^{k}P_{j,i}b_j \leq s_i  \leq \frac{b_i}{L_i}+\frac{1}{2}+\frac{1}{2L_i}\sum_{j=1}^{k}P_{j,i}b_j.
\end{eqnarray*}
This interval contains  only one integer number which is the unique solution
\begin{eqnarray}\label{si}
s_i=\left\lfloor \frac{1}{L_i} \left(b_i+ \frac{1}{2}\sum_{j=1}^{k}P_{j,i}b_j\right)\right\rceil .
\end{eqnarray}

Note that, after finding the shaped lattice codeword as discussed above, we must scale it by factor $2$ and then translate it by $(-1,\ldots , -1)$.
Then, the shaped vectors of $2\Lambda-(1,\ldots ,1)$ will be uniformly distributed over hypercube $2\mathcal{L}-(1,\ldots ,1)$, where
\begin{equation*}
\mathcal{L}=\left\{\mathbf{x}\in \mathbb{Z}^n\left|\begin{array}{ll}
                             \frac{-L_i}{2}\leq x_i \leq \frac{L_i}{2}-1, & i=1,\ldots,k \\
                              -L_i\leq x_i \leq L_i, & i=k+1,\ldots,n
                           \end{array}\right. \right\}.
\end{equation*}
Since, the lattice $2\Lambda$ and shaping region $\mathcal{L}$ are translated by the same vector and $\textrm{vol}(2\Lambda)=4^{n-k}2^k$, based on~(\ref{rate_loeliger}), the rate of our scheme is
\begin{equation}\label{rate}
  R=\frac{1}{n}\log_2\left(\frac{\textrm{vol}(\mathcal{L})}{\textrm{vol}(\Lambda)}\right) = \sum_{i=1}^k\frac{\log_2(L_i)}{n}+\sum_{i=k+1}^n\frac{\log_2\left(\frac{4L_i+2}{4}\right)}{n}.
\end{equation}

Algorithm~\ref{Unshaped} explains the method of obtaining original information $\mathbf{b}$ from shaped lattice codeword $\mathbf{x}'$.  The complexity of this algorithm is $O(nd)$, where $d$ is the average number of nonzero elements in a row of $\mathbf{G}_{\Lambda}$.
\begin{algorithm}
\small
 \begin{algorithmic}[1]
 \Procedure{MOD}{$\mathbf{x}',(L_1,\ldots,L_n),\mathbf{G}_{\Lambda}^{-1}$}
\State $\mathbf{b}'\gets\left\lfloor\left(\frac{\mathbf{x}'+1}{2}\right) \mathbf{G}_{\Lambda}^{-1}\right\rceil $
  \For{$i=1:n$} 
\If{$b_i' \pmod{L_i}< \frac{L_i}{2}$}
 \State $r_i \gets b_i' \pmod{L_i}$.
 \Else{}
  \State $r_i\gets b_i' \pmod{L_i}-L_i$
  \EndIf
   \EndFor
\State \textbf{return} $\mathbf{b}=(r_1,\ldots ,r_n)$.
 \EndProcedure
 \end{algorithmic}
 \caption{Obtaining original information}
 \label{Unshaped}
\end{algorithm}
\normalsize
\subsection{Nested Lattice Shaping Method}
Despite of its low complexity nature, hypercube shaping suffers a performance loss of $1.53$dB in high dimensions compared to optimal hypersphere shaping \cite{720542}. Thus, we consider nested lattice shaping, which is suboptimal but it offers more shaping gains comparing to hypercube shaping~\cite{5351439}.
First, limit the rate of the code by restricting the integer row  vector  $\mathbf{b}$ to take values  from a finite constellation in which $b_i\in \mathcal{L}_i=\left\{0,\ldots , L_i-1 \right\}$  for each $i=1,\ldots, n$. Similar to the hypercube shaping, let $\mathbf{x}'=(\mathbf{b}-\mathbf{sL})\mathbf{G}_{\Lambda}$. In this case, we choose the vector $\mathbf{s}$ as follows
\begin{eqnarray}\label{nested_lattice}
  \mathbf{s} =  \underset{\mathbf{s}_0\in\mathbb{Z}^n}{\mathrm{argmin}}\,\|(\mathbf{b}-\mathbf{s}_0\mathbf{L})\mathbf{G}_{\Lambda}\|^2.
\end{eqnarray}
Choosing $\mathbf{s}$ that minimizes $\|\mathbf{x}'\|$ is equivalent to  finding the nearest lattice point of the scaled lattice $\mathbf{LG}$ to the non-shaped lattice point $\mathbf{x}$. Therefore, the codewords will be uniformly distributed along the Voronoi cell of the coarse lattice $\mathbf{LG}$. Thus, the rate of the code is
\begin{equation}\label{rate_nested}
  R=\frac{1}{n}\log_2\left(\frac{\textrm{vol}(\mathbf{L}\Lambda)}{\textrm{vol}(\Lambda)}\right)=\frac{\sum_{i=1}^n\log_2(L_i)}{n}.
\end{equation}
The complexity of solving~(\ref{nested_lattice}) is exponential in the dimension of lattice, even by restricting the components of $\mathbf{b}$. Using the triangular structure of the parity check matrix $\mathbf{H}$, the authors of~\cite{5351439} suggested a tree search with affordable complexity for shaping their lattices.
Practically, their tree search can be done with simple sub-optimal sequential
decoders such as the $M$-algorithm~\cite{M_alg}. Following their method, we present a nested lattice shaping by using the triangular structure of our generator matrix.

Our algorithm starts from the first row of $\mathbf{G}_{\Lambda}$, and sequentially goes down along
the rows. The input at row $i$ is a list of up to $M$ candidate sequences $\left\{\mathbf{s}_1^{i-1},\ldots,\mathbf{s}_M^{i-1}\right\}$ for $\mathbf{s}_{i-1}=(s_1,\ldots,s_{i-1})$. For $i = 1$ the list is initialized with a single empty sequence. Each of these $M$ sequences will be extended to $(\mathbf{s}_j^{i-1},s')$, for $j=1,\ldots,M$, where $s'$ takes all possible values for $s_i$. The algorithm assigns to each extended sequence a respective score $\sum_{j=1}^i|x_j'|^2$. The scores are
sorted, and the $M$ sequences with smallest score are retained as input to the next row.  Finally, we choose $\mathbf{s}$ as the sequence with smallest score after processing of the last row $i = n$. The storage and the computational complexity of this algorithm are determined by the value of $M$, as $O(ndM)$. Using an $M$-algorithm with $M=1$, nested lattice shaping is equivalent to hypercube shaping, where for $M = \infty$, the algorithm is a full exponential tree search that finds the exact solution for $\mathbf{s}$.
The shaping gain and shaping loss of  LDPC lattice codes are presented in~\tablename{~\ref{table1}}. The results are obtained by using Monte Carlo simulation and $L_1=L_2=\cdots=L_n=L$ assumed to be $4$. An $M$-algorithm with $M=5$ has been applied for nested lattice shaping. The results show that increasing the rate of the code, used as underlying code of the lattice, declines the shaping gain for hypercube shaping but it improves the shaping gain for nested lattice shaping. Increasing the dimension of the lattice also improves the shaping gain for both of the applied shaping methods. Thus, the reasonable shaping gains can be obtained by applying nested lattice shaping for high dimensional  LDPC lattices, which have  high rate LDPC codes as their underlying codes.
\begin{table}[h]
\centering
\caption{The shaping gain and shaping loss for LDPC lattice codes in small dimensions.}
\renewcommand{\arraystretch}{1.1}
\small
\label{table1}
\begin{tabular}{cc|c|c||c|c|}
\cline{3-6}
\multicolumn{2}{c}{}                & \multicolumn{2}{|c||}{Hypercube shaping} & \multicolumn{2}{c|}{Nested lattice shaping} \\  \hhline{----||--}
\multicolumn{1}{|c|}{$(n,k)$}     & $L$ &         Shaping gain           &       Shaping loss            &    Shaping gain                   &  Shaping loss                    \\ \hhline{====::==}
\multicolumn{1}{|c|}{$(100,70)$}  &  $4$ & $-1.3065$ & $2.6756$   &  $-1.0162$   &  $2.3853$   \\
\multicolumn{1}{|c|}{$(100,80)$}  &  $4$ & $-1.4842$ & $2.8533$   &  $-0.7846$   &  $2.1536$   \\
\multicolumn{1}{|c|}{$(100,90)$}  &  $4$ & $-1.6665$ & $3.0356$   &  $-0.3717$   &  $1.7408$   \\
\multicolumn{1}{|c|}{$(200,100)$} & $4$  & $-0.9291$ & $2.3653$   &  $-0.8750$   &  $2.3112$   \\
\multicolumn{1}{|c|}{$(200,120)$} & $4$  & $-1.1124$ & $2.5486$   &  $-0.8926$   &  $2.3288$   \\
\multicolumn{1}{|c|}{$(200,140)$} & $4$  & $-1.2968$ & $2.7330$   &  $-0.8395$   &  $2.2757$   \\
\multicolumn{1}{|c|}{$(200,160)$} & $4$  & $-1.4798$ & $2.9160$   &  $-0.6593$   &  $2.0955$   \\
\multicolumn{1}{|c|}{$(200,180)$} & $4$  & $-1.6621$ & $3.0983$   &  $-0.3203$   &  $1.7564$   \\
\multicolumn{1}{|c|}{$(200,190)$} & $4$  & $-1.7565$ & $3.1927$   &  $-0.0633$   &  $1.4995$   \\ \hhline{----||--}
\end{tabular}
\end{table}
\section{One-Way Relay Network}\label{one_way_channel}
In this section,  we construct a lattice-coding scheme base on  LDPC lattice codes that achieves the decode-and-forward bound of (\ref{rate_bound}). We present our scheme by following the scheme of \cite{DBLP:conf/icc/FerdinandNA13} for LDLCs. All steps of this scheme are rephrased due to the  inherent differences between  LDPC lattice codes and LDLCs.
\subsection{Decomposition of  LDPC Lattice Codebooks}
Based on block Markov encoding and decomposition of the full lattice codebook into lower-rate codebooks, we can propose an encoding/decoding scheme for implementing  LDPC lattice codes on one-way relay channels. Let
$\mathbf{b}\in \mathbb{Z}^n$ be the information vector, where each element $b_i$ is drawn from the finite constellation of (\ref{constelation}).  Then, we define the $i^{th}$ element of the \emph{resolution component} $\mathbf{b}_r$ as
\begin{eqnarray}\label{constelation_r}
  b_{r,i} = \left\{ \begin{array}{l}
                        b_i,\quad i\in \mathcal{X}\\
                        0,\quad\, i\in \left\{1,\ldots ,n \right\}\backslash\mathcal{X},
                      \end{array}
  \right.
\end{eqnarray}
where $\mathcal{X}$ is a $k$-element  subset of $\left\{ 1,\ldots ,n \right\}$, whose members are chosen randomly. Let $\mathbf{b}_v \in \mathbb{Z}^n$ be the \emph{vestigial component} vector. Define the the $i^{th}$ element of $\mathbf{b}_v$ as follows
\begin{eqnarray}\label{constelation_v}
  b_{v,i} = \left\{ \begin{array}{l}
                        0,\quad\, i\in \mathcal{X}\\
                        b_i,\quad i\in \left\{1,\ldots ,n \right\}\backslash\mathcal{X}.
                      \end{array}
  \right.
\end{eqnarray}
Based on (\ref{constelation_r}) and (\ref{constelation_v}), define the resolution lattice codeword $\mathbf{x}_r=\mathcal{E}(b_r)$ and the vestigial lattice codeword $\mathbf{x}_v=\mathcal{E}(b_v)$, where $\mathcal{E}$ is defined by (\ref{encoding}). It can be easily checked that the original lattice codeword $\mathbf{x} = \mathcal{E}(\mathbf{b})$ is the sum of its resolution and vestigial components, that is
\begin{equation}\label{decomposition}
 \mathbf{x}=\mathbf{x}_r\oplus \mathbf{x}_v=\mathbf{x}_r+\mathbf{x}_v+(1,\ldots ,1).
\end{equation}
\subsection{ Power-Constrained Decomposition of  LDPC Lattices}
In preceding subsection, we used unconstrained powers for codewords $\mathbf{x}$, $\mathbf{x}_r$ and $\mathbf{x}_v$,  due to the fact that we still have not enforced shaping for the lattice. Linear decomposition is straightforward  in the unconstrained power situation, but it is not trivial with  power constrained scenario. We use the altered version of the proposed method in \cite{DBLP:conf/icc/FerdinandNA13} with the proposed  hypercube shaping method of Section~\ref{LDPC Lattices}. The steps of this method are as follows:
\begin{enumerate}
\item Map the information integer vector $\mathbf{b}$ to $\mathbf{b}'$ such that the $i^{th}$ element of $\mathbf{b}'$ is $b_i'=b_i-s_iL_i$ and
\begin{equation}\label{si2}
s_i=\left\{ \begin{array}{c}
              \left\lfloor \frac{1}{L_i} \left(b_i+ \frac{1}{2}\sum_{j=1}^{k}P_{j,i}b_j\right)\right\rceil ,\,\,\,\, i=k+1,\ldots ,n  \\
              0,\quad\quad\quad\quad\quad\quad\quad\quad\quad\quad  i=1,\ldots ,k.
            \end{array}
\right.
\end{equation}
The power-constrained lattice codeword is then given as $\mathbf{x}'=\mathcal{E}(\mathbf{b}')$.
  \item  Decompose the original integer information vector $\mathbf{b}$ as in (\ref{constelation_r}) and (\ref{constelation_v}).
  \item Map the resolution component $\mathbf{b}_r$ to $\mathbf{b}_r'$ such that the new integer vector results in a power-constrained codeword:
      \begin{equation}\label{bri}
b_{r,i}'=b_{r,i}-s_{r,i}L_i=\left\{ \begin{array}{l}
                        b_i-s_{r,i}L_i,\quad i\in \mathcal{X},\\
                        -s_{r,i}L_i,\quad\quad\, i \notin \mathcal{X},
                      \end{array}
\right.
\end{equation}
where $b_{r,i}$ and $b_{r,i}'$ are the $i^{th}$ elements of $\mathbf{b}_r$ and $\mathbf{b}_r'$, respectively. For hypercube shaping, $s_{r,i}$ can be written as,
      \begin{equation}\label{sri}
s_{r,i}=\left\{ \begin{array}{l}
                        \left\lfloor \frac{1}{L_i} \left(b_i+ \frac{1}{2}\sum_{j=1}^{k}P_{j,i}b_{r,j}\right)\right\rceil ,\quad\quad i\in \mathcal{I}\cap \mathcal{X}  \\
                        \\
\left\lfloor \frac{1}{2L_i} \left( \sum_{j=1}^{k}P_{j,i}b_{r,j}\right)\right\rceil ,\quad\quad\,\,\,\quad\quad i\in \mathcal{I}\cap \mathcal{X}^c \\
\\
                        0,\quad\quad\quad\quad\quad\quad\quad\quad\quad\quad\quad\quad\,\,\,\, i=1,\ldots ,k,
                      \end{array}
\right.
\end{equation}
where $\mathcal{I}=\left\{k+1,\ldots ,n\right\}$. Then, the mapped lattice codeword is $\mathbf{x}_r'=\mathcal{E}(\mathbf{b}_r')$.
  \item For the sake of preserving the linearity of the lattice decomposition, we map the vestigial information vector $\mathbf{b}_v$ to $\mathbf{b}_v'$ such that the $i_{th}$ element of $\mathbf{b}_v'$ is given as,
\begin{eqnarray}\label{bvi}
b_{v,i}'&=&b_i'-b_{r,i}'=(b_i-b_{r,i})+(s_{r,i}-s_i)L_i \nonumber\\
&=&\left\{ \begin{array}{l}
(s_{r,i}-s_i)L_i,\quad\quad\quad i\in \mathcal{X}\\
b_i+(s_{r,i}-s_i)L_i,\quad i \notin \mathcal{X},
\end{array}
\right.
\end{eqnarray}
where $s_i$ and $s_{r,i}$ are given in (\ref{si2}) and (\ref{sri}), respectively. Then, the vestigial codeword $\mathbf{x}_v'$ can be obtained as $\mathbf{x}_v'=\mathcal{E}(\mathbf{b}_v')$.
\end{enumerate}

The primary information integers $b_i$, $b_{r,i}$ and $b_{v,i}$ can be recovered from
$b_i'$, $b_{r,i}$ and $b_{v,i}$ by modulo $L_i$ operation. It must be noted that  the lattice codeword $\mathbf{x}'$ and the resolution lattice codeword $\mathbf{x}_r'$ respect the power constraint but, the vestigial lattice codeword $\mathbf{x}_v'$ may not fulfil the power constraint in general. However, this is not a problem for the considered relay network since we do not transmit the vestigial information alone. In order to obtain further shaping gain, we employ nested lattice shaping in our decomposition. In this case, we find the vectors $\mathbf{s}$ and $\mathbf{s}_r$
such that
\begin{eqnarray}
  \mathbf{s}&=&  \underset{\mathbf{s}^{*}\in\mathbb{Z}^n}{\mathrm{argmin}}\,\|(\mathbf{b}-\mathbf{s}^{*}\mathbf{L})\mathbf{G}_{\Lambda}\|^2,\label{sr_s_nested}\\
 \mathbf{s}_r&=&  \underset{\mathbf{s}_r^{*}\in\mathbb{Z}^n}{\mathrm{argmin}}\,\|(\mathbf{b}_r-\mathbf{s}_r^{*}\mathbf{L})\mathbf{G}_{\Lambda}\|^2.\label{sr_s_nested2}
\end{eqnarray}
After finding $\mathbf{s}$ and $\mathbf{s}_r$, the vestigial codeword can be found as $\mathbf{x}_v'=\mathcal{E}\left(\mathbf{b_v}+(\mathbf{s}-\mathbf{s}_r)\mathbf{L}\right)$.
\subsection{Encoding and Decoding}\label{encoding_oneway}
Here, we explain the implementation of the encoding and the decoding of  LDPC lattice codes, for  one-way relay network.
Let $\mathbf{x}'[t]$ be the full-rate power-constrained lattice codeword associated with $t^{th}$ block, and let $\mathbf{x}_r'[t-1]$ be the
resolution codeword decoded at the relay from the $[t-1]^{th}$ block. The source
and the relay transmit $\sqrt{P_S}\mathbf{x}'[t]$ and $\sqrt{P_R}\mathbf{x}_r'[t-1]$ during the
$t^{th}$ block. At block $t=1$, the source sends $\sqrt{P_S}\mathbf{x}'[1]$ and the relay sends nothing. At blocks $t=2,\ldots , T$, destination receives the superposition of signals from the source and the relay at blocks $t$ and $t-1$, respectively. At block $t=T+1$, destination only receives the resolution information from the relay since the source has no information to send. The source transmits $nTR$ symbols over $n(T+1)$ channel use, so the encoding rate is close to $R$ for large $T$. More detail about implementing block Markov encoding using lattices can be found in \cite{6506106,DBLP:conf/icc/FerdinandNA13} and \cite{6994267}.

Following the outline that proposed in \cite[Theorem 1]{6994267}, our decoding contains three stages. At the first stage, the relay decodes $\mathbf{x}'$, in the next stage, the destination decodes $\mathbf{x}_r'$, finally, the destination uses $\mathbf{x}_r'$ to decode $\mathbf{x}_v'$.
First, we consider the decoding at the relay.  The received signal by the relay at the $t^{th}$ block  is
\begin{equation}\label{yr}
\mathbf{y}_R[t]=h_{SR}\sqrt{P_S}\mathbf{x}'[t]+\mathbf{z}_R[t].
\end{equation}
We start by scaling $\mathbf{y}_R[t]$ by the factor $1/\sqrt{P_S}h_{SR}$. Then, we use iterative decoder of  LDPC lattices to obtain full-rate lattice codeword $\hat{\mathbf{x}}'[t]$
\begin{equation}\label{stage1}
\hat{\mathbf{x}}'[t]=\textrm{DEC}_{\textrm{LDPCL}}\left( \frac{\mathbf{y}_R[t]}{\sqrt{P_S}h_{SR}},\frac{N_R}{\sqrt{P_S}h_{SR}} \right),
\end{equation}
where $\textrm{DEC}_{\textrm{LDPCL}}(\mathbf{y},\sigma^2)$ is the iterative decoding algorithm of  LDPC lattices described in Section~\ref{LDPC Lattices}. We use $\mathbf{y}$  and $\sigma^2$ to estimate LLR vector in (\ref{LLR}). The resulting information vector is denoted by $\hat{\mathbf{b}}[t]$ and can be obtained as $\hat{\mathbf{b}}[t]=\textrm{MOD}(\hat{\mathbf{x}}'[t],\mathbf{L},\mathbf{G}_{\Lambda}^{-1})$, where $\mathbf{G}_{\Lambda}^{-1}$ is
\begin{eqnarray}\label{eq122}
  \mathbf{G}_{\Lambda}^{-1}&=& \left[
                    \begin{array}{cc}
                      \frac{1}{2}\mathbf{I}_{k}& -\frac{1}{4}\mathbf{P}_{k\times (n-k)} \\
                      \mathbf{0}_{(n-k)\times k} & \frac{1}{4}\mathbf{I}_{n-k} \\
                    \end{array}
                  \right].
\end{eqnarray}
The relay estimates the resolution information $\hat{\mathbf{b}}_r[t]$ from  $\hat{\mathbf{b}}[t]$ and (\ref{constelation_r}), from which it finds the estimation of shaped resolution codeword  $\hat{\mathbf{x}}_r'[t]$. In this case, the  decoding error is $\mathbf{e}_r[t]=\mathbf{x}_r'[t]-\hat{\mathbf{x}}_r'[t]$. The received signal at the destination during $(t + 1)^{th}$ block is
\begin{eqnarray*}
\mathbf{y}_D[t+1]=h_{RD}\sqrt{P_R}\hat{\mathbf{x}}_r'[t]+h_{SD}\sqrt{P_S}\mathbf{x}'[t+1]+\mathbf{z}_D[t+1].
\end{eqnarray*}
Replace $\hat{\mathbf{x}}_r'[t]$ by $\mathbf{x}_r'[t]-\mathbf{e}_r[t]$ in $\mathbf{y}_D[t+1]$. In the destination, treat $h_{SD}\sqrt{P_S}\mathbf{x}'[t+1]+\mathbf{z}_D[t+1]-h_{RD}\sqrt{P_R}\mathbf{e}_r[t]$  as noise and decode the resolution information vector $\tilde{\mathbf{x}}_r'[t]$
\begin{eqnarray*}
\tilde{\mathbf{b}}_r'[t]=\left\lfloor \mathcal{D}\left(\textrm{DEC}_{\textrm{LDPCL}}\left( \frac{\mathbf{y}_D[t+1]}{\sqrt{P_R}h_{RD}},\frac{N_D}{\sqrt{P_R}h_{RD}} \right)\right)\mathbf{G}_{\Lambda}^{-1}\right\rceil,
\end{eqnarray*}
where $\mathcal{D}(x_1,\ldots ,x_n)=0.5\times(x_1+1,\ldots,x_n+1)$.

Then, the relay  obtains  $\tilde{\mathbf{b}}_r[t]=\textrm{MOD}(\tilde{\mathbf{x}}_r'[t],\mathbf{L},\mathbf{G}_{\Lambda}^{-1})$, from which it finds the unshaped resolution codeword $\tilde{\mathbf{x}}_r[t]$. The decoding errors are
\begin{eqnarray}
  \mathbf{e}_{d_1}'[t] &=&\mathbf{x}_r'[t]- \tilde{\mathbf{x}}_r'[t], \nonumber  \\
  \mathbf{e}_{d_1}[t] &=& \mathbf{x}_r[t] -\tilde{\mathbf{x}}_r[t].
\end{eqnarray}
Now, the destination knows the unshaped resolution codeword  $\tilde{\mathbf{x}}_r[t]$ and the shaped resolution codeword $\tilde{\mathbf{x}}_r'[t]$. Next, the receiver turns to $\mathbf{y}_D[t]$, which can be written as
\begin{eqnarray}\label{y_Dt}
\mathbf{y}_D[t]&=&h_{RD}\sqrt{P_R}(\mathbf{x}_r'[t-1]-\mathbf{e}_r[t-1])+ \nonumber\\
 &&\quad\quad\quad\quad\quad\quad\quad h_{SD}\sqrt{P_S}\mathbf{x}'[t]+\mathbf{z}_D[t].
\end{eqnarray}
From $\mathbf{y}_D[t]$, we know $\tilde{\mathbf{x}}_r'[t-1]$. By using the linearity property~(\ref{decomposition}), we can rewrite (\ref{y_Dt}) as follows
\begin{eqnarray}\label{y_Dt2}
\mathbf{y}_D[t]&=&h_{RD}\sqrt{P_R}(\mathbf{x}_r'[t-1]-\mathbf{e}_r[t-1])+ \mathbf{z}_D[t]+\nonumber\\
 && h_{SD}\sqrt{P_S}(\mathbf{x}_r'[t]+\mathbf{x}_v'[t]+(1,\ldots ,1)).
\end{eqnarray}
Now, we subtract the decoded resolution information  $h_{SD}\sqrt{P_S}(\tilde{\mathbf{x}}_r'[t]+(1,\ldots ,1))+h_{RD}\sqrt{P_R}\tilde{\mathbf{x}}_r'[t-1]$ from (\ref{y_Dt2}) to obtain
\begin{eqnarray}\label{y_Dt3}
\mathbf{y}_D'[t]=h_{SD}\sqrt{P_S}\mathbf{x}_v'[t]+\mathbf{e}_{d_2}[t]+\mathbf{z}_D[t],
\end{eqnarray}
where
\begin{eqnarray*}
\mathbf{e}_{d_2}[t]=h_{RD}\sqrt{P_R}(\mathbf{e}_{d_1}'[t-1]-\mathbf{e}_r[t-1])+ h_{SD}\sqrt{P_S}\mathbf{e}_{d_1}'[t].
\end{eqnarray*}
Then, we use $\mathbf{y}_D'[t]$ in (\ref{y_Dt3}) to decode the vestigial information as follows
\begin{eqnarray}
\tilde{\mathbf{x}}_v'[t]=\textrm{DEC}_{\textrm{LDPCL}}\left( \frac{y_D'[t]}{\sqrt{P_S}h_{SD}},\frac{N_D}{\sqrt{P_S}h_{SD}} \right).
\end{eqnarray}
Once we have decoded both the resolution and vestigial lattice codeword, the destination can find the desired lattice codeword by $\tilde{\mathbf{x}}'[t]=\tilde{\mathbf{x}}_v'[t]\oplus\tilde{\mathbf{x}}_r'[t]=\tilde{\mathbf{x}}_v'[t]+\tilde{\mathbf{x}}_r'[t]+(1,\ldots,1)$. Then, the desired  integer vector can be
obtained as $\tilde{\mathbf{b}}[t]= \textrm{MOD}(\tilde{\mathbf{x}}'[t],\mathbf{L},\mathbf{G}_{\Lambda}^{-1})$.
\section{Two-Way Relay Networks}\label{two_way_channel_sec}
In this section,  we present a practical block Markov encoding scheme for the
two-way relay channel based on  LDPC lattice codes. This scheme  achieves the decode-and-forward bound in  (\ref{rate_bound2}) and it is proposed in \cite{6994267}  for LDLCs. Due to the structural differences between  LDPC lattices and LDLCs, as well as, differences between our shaping methods and theirs, all steps of this scheme are rephrased for LDPC lattice codes.

Our applied scheme relies on doubly-nested lattice codes in which the two sources use different codebooks with different transmit powers.
Let $\Lambda_{S_1}^{s}\subset \Lambda_{S_1}^{c}$ and $\Lambda_{S_2}^{s}\subset \Lambda_{S_2}^{c}$ be the shaping and coding lattices for $S_1$ and $S_2$, respectively.  Consider the codebooks $\mathcal{C}_{S_1} =\Lambda_{S_1}^{c}\cap \mathcal{V}(\Lambda_{S_1}^{s})$  and $\mathcal{C}_{S_2} =\Lambda_{S_2}^{c}\cap \mathcal{V}(\Lambda_{S_2}^{s})$.
It is also assumed by the authors of~\cite{6994267} that $h_{S_1R}P_{S_1}\geq h_{S_2R}P_{S_2}$, $\sigma^2(\Lambda_{S_1}^{s})=h_{S_1R}P_{S_1}$, and $\sigma^2(\Lambda_{S_2}^{s})=h_{S_2R}P_{S_2}$.
There is another lattice, which is referred to as  \emph{meso-lattice}, that  partitions the lattice codebook into lower-rate constituent codebooks \cite{6994267}.
The meso-lattice $\Lambda_m$, is nested according to $\Lambda_{S_1}^{s}\subset \Lambda_{S_2}^{s}\subset \Lambda_m \subset \Lambda_{S_1}^{c},\Lambda_{S_2}^{c}$, where $\Lambda_{S_1}^{c}$ and $\Lambda_{S_2}^{c}$ are also nested
lattices. Define the resolution codebooks $\mathcal{C}_{S_1}^{(r)} =\Lambda_{S_1}^{c}\cap \mathcal{V}(\Lambda_m)$  and $\mathcal{C}_{S_2}^{(r)} =\Lambda_{S_2}^{c}\cap \mathcal{V}(\Lambda_m)$. Then, the vestigial codebooks are $\mathcal{C}_{S_1}^{(v)} = \Lambda_m \cap \mathcal{V}(\Lambda_{S_1}^{s})$  and $\mathcal{C}_{S_2}^{(v)} = \Lambda_m \cap \mathcal{V}(\Lambda_{S_2}^{s})$.
Let $R_{S_1},R_{S_2},R_{S_1}^{(r)},R_{S_2}^{(r)},R_{S_1}^{(v)},R_{S_2}^{(v)}$ be the rates of $\mathcal{C}_{S_1},\mathcal{C}_{S_2},\mathcal{C}_{S_1}^{(r)},\mathcal{C}_{S_2}^{(r)},\mathcal{C}_{S_1}^{(v)},\mathcal{C}_{S_2}^{(v)}$, respectively. Then, as before, $R_{S_1}=R_{S_1}^{(r)}+R_{S_1}^{(v)}$ and $R_{S_2}=R_{S_2}^{(r)}+R_{S_2}^{(v)}$. Thus, each full-rate codeword can be decomposed into a unique modulo sum of resolution and vestigial
codewords. It is proved that the preceding code construction achieves near-capacity rates~\cite{6994267}.
\subsection{Decomposition of LDPC Lattice Codebooks}
Let $\mathbf{b}\in \mathbb{Z}^n$ be the information vector, where each element $b_i$ of $\mathbf{b}$ is drawn from  finite constellations $\left\{-L_i/2,\ldots,(L_i/2) -1\right\}$ and $\left\{0,\ldots,L_i-1\right\}$ for hypercube shaping and nested lattice shaping, respectively. The $i^{th}$ element of the resolution component
$\mathbf{b}_r$ is
\begin{eqnarray}\label{resolution_twoway}
b_{i}^{(r)}=b_i\pmod{L_{i}^{(r)}},
\end{eqnarray}
where $L_i=\beta L_{i}^{(r)}$, for some $\beta\in \mathbb{Z}$. Thus, the $i^{th}$ element of $\mathbf{b}^{(r)}$  lies in the finite constellation $\mathcal{L}_i^{(r)}=\left\{0,\ldots,L_{i}^{(r)}-1\right\}$ and the rate of the obtained codebook $R^{(r)}$ can be acquired via~(\ref{rate}) or (\ref{rate_nested}), depending on the applied shaping method, by replacing $L_i$'s with $L_i^{(r)}$'s.
The vestigial component is defined as follows
\begin{equation}\label{vestigial_two_way}
  \mathbf{b}^{(v)}=\mathbf{b}-\mathbf{b}^{(r)}.
\end{equation}
If the employed shaping method is hypercube shaping, it can be shown that the $i^{th}$ element of $\mathbf{b}^{(v)}$ lies in the following finite constellation
\begin{equation}\label{vestigial_constelation}
  \mathcal{L}_i^{(v)}=\left\{\frac{-\beta }{2}L_i^{(r)},\frac{(-\beta+2) }{2}L_i^{(r)},\ldots,\frac{(\beta-2) }{2}L_i^{(r)}\right\}.
\end{equation}
If the employed shaping method is nested lattice shaping, the $i^{th}$ element of $\mathbf{b}^{(v)}$ lies in the finite constellation $\left\{0,L_i^{(r)},2L_i^{(r)},\ldots,(\beta-1)L_i^{(r)}\right\}$.
The rate of the vestigial  codebook is approximately $R^{(v)}=\log_2(\beta)$. Due to the aforementioned assumptions, each codeword $\mathbf{x}=\mathcal{E}(\mathbf{b})$ decomposes into  its resolution component $\mathbf{x}^{(r)}=\mathcal{E}(\mathbf{b}^{(r)})$ and its vestigial component $\mathbf{x}^{(v)}=\mathcal{E}(\mathbf{b}^{(v)})$ as $\mathbf{x}=\mathbf{x}^{(r)}\oplus \mathbf{x}^{(v)}=\mathbf{x}^{(r)}+ \mathbf{x}^{(v)}+(1,\ldots,1)$.
\subsection{Power-Constrained Decomposition of LDPC Lattice Codebooks}
The first step in shaping the lattice codeword $\mathbf{x}$ is translating $b_i$ by an integer multiple of $L_i$. Then, following the same procedure as in Section~\ref{shaping_sec}, the shaped lattice point is obtained.  The shaped lattice codeword is $\mathbf{x}'=\mathcal{E}(\mathbf{b}-\mathbf{sL})$, where $\mathbf{s}$  is given in (\ref{si}) and (\ref{nested_lattice}), for hypercube shaping and nested lattice shaping, respectively. Next, we adapt the shaping methods to the resolution codewords in such a way that the decomposition of the lattice codebook remains linear. We shape the resolution component $\mathbf{b}^{(r)}$ to $\mathbf{b}^{\prime(r)}$ as follows
\begin{equation}\label{b_r_prime_twoway}
 b_i^{\prime(r)}=\mathrm{smod}(b_i,L_i^{(r)})-s_i^{(r)}L_i^{(r)},
\end{equation}
where $b_i$ and  $b_i^{\prime(r)}$ are the $i^{th}$ elements of $\mathbf{b}$ and $\mathbf{b}^{\prime(r)}$, respectively, and
\begin{equation}\label{smod_func}
  \mathrm{smod}(x,L)=\left\{\begin{array}{l}
                       \bar{x}=x \,\,(\bmod \,\,L),\quad  \mathrm{if}\,\, \bar{x}<\frac{L}{2},  \\
                       x \,\,(\bmod \,\,L)-L,\quad \mathrm{otherwise}.
                     \end{array}\right.
\end{equation}
We choose the elements of $\mathbf{s}^{(r)}$ according to
 \begin{equation}\label{s_r_twoway}
   s_i^{(r)}=\left\lfloor\frac{1}{L_i^{(r)}}\left(b_i^{(r)}+\frac{1}{2}\sum_{j=1}^kP_{j,i}b_j^{(r)}\right)\right\rceil,\quad i=1,\ldots,n,
 \end{equation}
for hypercube shaping. For nested lattice shaping, we consider $b_i^{\prime(r)}=b_i\pmod{L_i^{(r)}}-s_i^{(r)}L_i^{(r)}$, where $\mathbf{s}^{(r)}$ is given by (\ref{sr_s_nested2}). Indeed, the $\mathrm{smod}$ function is regular modulo operation, when the  employed shaping method is nested lattice shaping. Then, the shaped resolution component  is given by $\mathbf{x}^{\prime(r)}=\mathcal{E}(\mathbf{b}^{\prime(r)})$. In order to preserve the modulo linearity of the lattice decomposition, we map the vestigial integer vector $\mathbf{b}^{(v)}$ to $\mathbf{b}^{\prime(v)}$ as follows
\begin{equation}\label{b_v_prime_twoway}
  \mathbf{b}^{\prime(v)}=\mathbf{b}'-\mathbf{b}^{(r)}=\mathbf{b}^{(v)}-\mathbf{sL},
\end{equation}
where $\mathbf{s}$  is given in (\ref{si}) and (\ref{nested_lattice}), for hypercube shaping and nested lattice shaping, respectively. According to this decomposition, we have $\mathbf{b}'\neq \mathbf{b}^{\prime(r)}+\mathbf{b}^{\prime(v)}$ in general. However, the decomposition preserves componentwise modulo linearity, that is, $\mathbf{b}_i'\pmod{L_i}= \mathbf{b}_i^{\prime(r)}\pmod{L_i^{(r)}}+\mathbf{b}_i^{\prime(v)}\pmod{L_i}$.
Then, the  vestigial codeword is $\mathbf{x}^{\prime(v)}=\mathcal{E}(\mathbf{b}^{\prime(v)})$. It can be easily shown that $\mathbf{x}'=\mathbf{x}^{(r)}\oplus \mathbf{x}^{\prime(v)}$. It should be noted that  the vestigial lattice codeword $\mathbf{x}_v'$ may not fulfil the power constraint, which is not a problem for the considered relay network, since we do not transmit the vestigial information alone. The original information vectors $\mathbf{b},\mathbf{b}^{(v)}$ can be recovered from $\mathbf{b}',\mathbf{b}^{\prime(v)}$ by using Algorithm~\ref{Unshaped} with $\mathbf{L}=(L_1,\ldots,L_n)$. Similarly,
the resolution information vector $\mathbf{b}^{(r)}$ can be recovered from $\mathbf{b}^{\prime(r)}$ by using Algorithm~\ref{Unshaped} with $\mathbf{L}^{(r)}=(L_1^{(r)},\ldots,L_n^{(r)})$.
\subsection{Encoding and Decoding}
Here, we present the implementation of encoding and decoding for two-way relay network using LDPC
lattice codes. First, we describe the encoding scheme and then we discuss the decoding schemes at each node.
The two sources employ LDPC lattice codes as described in previous subsection.
The full-rate information vector, resolution information vector,
vestigial information vector, full-rate codeword, resolution
codeword, and vestigial codeword of source $i$ are denoted by $\mathbf{b}_{S_i},\mathbf{b}_{S_i}^{(r)},\mathbf{b}_{S_i}^{(v)},\mathbf{x}_{S_i},\mathbf{x}_{S_i}^{(r)}$ and $\mathbf{x}_{S_i}^{(v)}$, respectively, for $i=1, 2$.
Let the constellation sizes of the $i^{th}$ element of $\mathbf{b}_{S_1}$ and $\mathbf{b}_{S_2}$ be $L_{S_1,i}$ and $L_{S_2,i}$, respectively. Further, let $L_i^{(r)}$ be the constellation size of the resolution codeword, which is selected such that $L_i^{(r)}$ divides both $L_{S_1,i}$ and $L_{S_2,i}$ \cite{6994267}. It is assumed by the authors of \cite{6994267} that
the sources use transmit powers such that
\begin{equation}\label{powe_condition}
  \frac{\sqrt{P_{S_1}}h_{S_1R}}{m_1}=\frac{\sqrt{P_{S_2}}h_{S_2R}}{m_2}=\rho ,
\end{equation}
where $m_1,m_2\in \mathbb{Z}$ and $\rho\in \mathbb{R}$.  Furthermore, it is assumed that  $\gcd (L_i^{(r)},m_1)=\gcd (L_i^{(r)},m_2)=1$. In the rest of this paper, we assume $m_1=m_2=1$.

First, we describe the encoding steps.  The two sources $S_1$ and $S_2$ transmit their signals to the relay and to  each other. Meanwhile, the relay transmits its own signal to both $S_1$ and $S_2$. Thus, $S_1$ receives the superposition of signals from $S_2$ and the relay. Similarly, $S_2$ receives signals from $S_1$ and the relay. During the $t^{th}$ block, where $t=2,\ldots,T$, the sources $S_1$ and $S_2$ transmit their new codewords $\sqrt{P_{S_1}}\mathbf{x}_{S_1}'[t]$ and $\sqrt{P_{S_2}}\mathbf{x}_{S_2}'[t]$, while  the relay transmits the resolution component of the decoded sum codeword $\sqrt{P_R}\mathbf{x}_R^{\prime(r)}[t-1]$. During the block $t=1$, the relay transmits nothing and at block $t=T+1$, the  sources receive the resolution information $\sqrt{P_R}\mathbf{x}_R^{\prime(r)}[T]$ from the relay, since both sources have no information to send.

Next, we describe the decoding steps. Since, the decoding at $S_1$ and $S_2$ is similar,  we only consider decoding at $S_2$. Decoding occurs in three phases. In phase one, the relay decodes the sum codeword $\mathbf{x}_{S_1}'\oplus \mathbf{x}_{S_2}'$. In phase two, $S_2$ decodes the resolution codeword $\mathbf{x}_{S_1}^{(r)}$ by treating other codewords as noise. In the last phase, $S_2$ decodes the vestigial codeword $\mathbf{x}_{S_1}^{(v)}$.

The received signal at relay in $t^{th}$ block is
\begin{eqnarray}
  \mathbf{y}_R[t] &=& \sqrt{P_{S_1}}h_{S_1R}\mathbf{x}_{S_1}'[t]+\sqrt{P_{S_2}}h_{S_2R}\mathbf{x}_{S_2}'[t]+\mathbf{z}_R[t]\nonumber \\
   &=& \rho\left(m_1\mathbf{x}_{S_1}'[t]+m_2 \mathbf{x}_{S_2}'[t]\right)+\mathbf{z}_R[t]. \label{y_R_two_way}
\end{eqnarray}
Let $\mathbf{y}_R'[t]=\mathbf{y}_R[t]-\rho\left(-m_1-m_2+1\right)$. Using $\mathbf{y}_R'[t]$, the relay performs LDPC lattice decoding to obtain an estimate of the sum information $m_1\mathbf{b}_{S_1}'[t]+m_2 \mathbf{b}_{S_2}'[t]$
\begin{eqnarray}
  m_1\hat{\mathbf{x}}_{S_1}'[t]\oplus m_2 \hat{\mathbf{x}}_{S_2}'[t] &=& \textrm{DEC}_{\textrm{LDPCL}}\left(\frac{\mathbf{y}_R'[t]}{\rho},\frac{N_R}{\rho}\right), \quad\,\,\,\label{sum_information1}\\
  m_1\hat{\mathbf{b}}_{S_1}'[t]+m_2 \hat{\mathbf{b}}_{S_2}'[t] &=&\left\lfloor \mathcal{D}\left(   \hat{\mathbf{x}}_R'[t]  \right) \mathbf{G}_{\Lambda}^{-1}\right\rceil,\label{sum_information2}
\end{eqnarray}
where function $\mathcal{D}$ is defined in Section~\ref{encoding_oneway} and $\hat{\mathbf{x}}_R'[t]=m_1\hat{\mathbf{x}}_{S_1}'[t]+m_2 \hat{\mathbf{x}}_{S_2}'[t]$. The $i^{th}$ element of the sum is given by
\begin{eqnarray}
\hat{b}_{R,i}'&=&m_1\hat{b}_{S_1,i}'+m_2 \hat{b}_{S_2,i}'\\
   &=& m_1(\hat{b}_{S_1,i}-s_{1,i}L_{S_1,i})+m_2(\hat{b}_{S_2,i}-s_{2,i}L_{S_2,i}) \nonumber \\
   &=& m_1\hat{b}_{S_1,i}+m_2\hat{b}_{S_2,i}- m_2s_{2,i}L_{S_2,i}-m_1s_{1,i}L_{S_1,i}.\nonumber
\end{eqnarray}
Then, the relay takes the result modulo $L_i^{(r)}$ to find the modulo sum of the two resolution information vectors
\begin{eqnarray}
  \hat{b}_{R,i}^{(r)} &=& \mathrm{smod}\left(\left[ m_1\hat{b}_{S_1,i}+m_2 \hat{b}_{S_2,i}\right],L_i^{(r)}\right) \\
   &=& \mathrm{smod}\left( \left[ m_1\hat{b}_{S_1,i}'+m_2 \hat{b}_{S_2,i}'\right],L_i^{(r)} \right).
\end{eqnarray}
Since $L_i^{(r)}$ and $m_1$, $m_2$ were selected  to be co-primes, the individual codewords can be recovered from a
single codeword and the modulo sum. Due to the existent constraints on  transmit power, the relay employs shaping methods and maps $\hat{\mathbf{b}}_{R}^{(r)}[t]$ to another information vector $\hat{\mathbf{b}}_{R}^{\prime(r)}[t]$
\begin{eqnarray}
  \hat{\mathbf{b}}_{R}^{\prime(r)}[t] &=& \hat{\mathbf{b}}_{R}^{(r)}[t]-\mathbf{s}_R\mathbf{L}^{(r)},
\end{eqnarray}
where $\mathbf{s}_R$ is given in (\ref{s_r_twoway}) and (\ref{sr_s_nested2}) for hypercube shaping and nested lattice shaping, respectively. Then,  the shaped lattice codeword is $\hat{\mathbf{x}}_{R}^{\prime(r)}[t]=\mathcal{E}\left(\hat{\mathbf{b}}_{R}^{\prime(r)}[t]\right)$. During block $t+1$, the relay transmits $\sqrt{P_R}\hat{\mathbf{x}}_{R}^{\prime(r)}[t]$.

Now, we describe the decoding process at $S_2$. The received signal at $S_2$ in block $t+1$ is
\begin{eqnarray}
  \mathbf{y}_{S_2}^{\prime}[t+1] &=& h_{RS_2}\sqrt{P_R}\hat{\mathbf{x}}_{R}^{\prime(r)}[t]\nonumber\\
  &+&h_{S_1S_2}\sqrt{P_{S_1}}\mathbf{x}_{S_1}'[t+1]+\mathbf{z}_{S_2}[t+1].
\end{eqnarray}
Source $2$ subtracts its own scaled and translated codeword $m_2h_{RS_2}\sqrt{P_R}\left[\mathbf{x}_{S_2}[t]+(1,\ldots,1)\right]$ from the received signal to obtain
\begin{eqnarray}
  \mathbf{y}_{S_2}^{\prime\prime}[t+1] &=& h_{RS_2}\sqrt{P_R}\left[\hat{\mathbf{x}}_{R}^{\prime(r)}[t]-m_2\mathbf{x}_{S_2}[t]-(1,\ldots,1)\right]\nonumber\\
  &+&h_{S_1S_2}\sqrt{P_{S_1}}\mathbf{x}_{S_1}'[t+1]+\mathbf{z}_{S_2}[t+1].
\end{eqnarray}
Then, treating $h_{S_1S_2}\sqrt{P_{S_1}}\mathbf{x}_{S_1}'[t+1]+\mathbf{z}_{S_2}[t+1]$ as noise, $S_2$ uses  LDPC lattice decoding to obtain
\begin{equation}\label{dec_S2_twoway}
 \tilde{\mathbf{b}}_{S_1}^{\prime (r)}[t]=\left\lfloor \mathcal{D}\left( \textrm{DEC}_{\textrm{LDPCL}}\left(\frac{\mathbf{y}_{S_2}^{\prime\prime}[t+1]}{ \gamma},\frac{N_{S_2}}{ \gamma}\right) \right) \mathbf{G}_{\Lambda}^{-1}\right\rceil,
\end{equation}
where $\gamma=h_{RS_2}\sqrt{P_R}$. It must be noted that the decoded integer vector in (\ref{dec_S2_twoway}) is
\begin{equation}\label{dec_S2_twoway2}
   \tilde{\mathbf{b}}_{S_1}^{\prime (r)}[t]=\mathrm{smod}\left( \tilde{\mathbf{b}}_R^{(r)}[t],\mathbf{L}^{(r)} \right)-m_2\mathbf{b}_{S_2}[t]-\mathbf{s}_R\mathbf{L}^{(r)},
\end{equation}
where $\tilde{\mathbf{b}}_R^{(r)}[t]= m_1\tilde{\mathbf{b}}_{S_1}[t]+m_2 \tilde{\mathbf{b}}_{S_2}[t]$. Let $\mathbf{b}_1[t]=\mathrm{smod}\left(\left[\tilde{\mathbf{b}}_{S_1}^{\prime (r)}[t]+m_2\mathbf{b}_{S_2}[t]\right],\mathbf{L}^{(r)} \right)$ and $\mathbf{b}_2[t]=\mathrm{smod2mod}\left(\mathbf{b}_1[t],\mathbf{L}^{(r)} \right)$,
in which the function $\mathrm{smod2mod}(x,L)$, that is defined next, is applied componentwise
\begin{equation}\label{smod2mod}
  \mathrm{smod2mod}(x,L)=\left\{\begin{array}{l}
                           x,\quad\quad\,\, \mathrm{if}\,\, 0\leq x\leq \frac{L}{2}-1, \\
                           x+L,\,\,\, \mathrm{if}\,\, \frac{-L}{2}\leq x <0.
                         \end{array}\right.
\end{equation}
Source $2$ obtains the resolution information of source $1$ as
\begin{eqnarray}\label{b_S1(r)_prime}
  \left(\mathbf{b}_2[t] \right. &-& \left. m_2\mathbf{b}_{S_2}[t]\right) \pmod*{\mathbf{L}^{(r)}}\nonumber\\
  &=&\left[\left(m_1\tilde{\mathbf{b}}_{S_1}[t]+m_2 \tilde{\mathbf{b}}_{S_2}[t]\right)\pmod*{\mathbf{L}^{(r)}} \right. \nonumber\\
  &&- \left. m_2\mathbf{b}_{S_2}[t]-\mathbf{s}_R\mathbf{L}^{(r)}\right]\pmod*{\mathbf{L}^{(r)}}\nonumber\\
  &=& m_1\tilde{\mathbf{b}}_{S_1}[t]\pmod*{\mathbf{L}^{(r)}}.
\end{eqnarray}
Then, it recovers the $i^{th}$ element of unshaped resolution information $\tilde{\mathbf{b}}_{S_1}^{(r)}[t]$ by computing
\begin{eqnarray}
  \tilde{b}_{S_1,i}^{(r)}= \frac{ (m_1\tilde{b}_{S_1,i})\pmod*{L_{i}^{(r)}} +\delta L_{i}^{(r)} }{m_1},
\end{eqnarray}
where $\delta$ is the unique integer such that $\tilde{b}_{S_1,i}^{(r)}\in\left\{0,\ldots, L_{i}^{(r)}\right\}$. Such
a unique $\delta$ always exists since $L_{i}^{(r)}$ and $m_1$ are coprime \cite{6994267}.
Next, $S_2$ uses $\tilde{\mathbf{b}}_{S_1}^{(r)}[t]$ and $\tilde{\mathbf{b}}_{S_1}^{\prime (r)}[t]+m_2\mathbf{b}_{S_2}[t]$  to obtain $\tilde{\mathbf{x}}_{S_1}^{(r)}[t]=\mathcal{E}(\tilde{\mathbf{b}}_{S_1}^{(r)}[t])$ and  $\tilde{\mathbf{x}}_{R}^{\prime(r)}[t]=\mathcal{E}(\tilde{\mathbf{b}}_{S_1}^{\prime (r)}[t]+m_2\mathbf{b}_{S_2}[t])$. Hence, it
has $\tilde{\mathbf{x}}_{R}^{\prime(r)}[t-1]$ and $\tilde{\mathbf{x}}_{S_1}^{(r)}[t]$ from $\mathbf{y}_{S_2}'[t]$ and $\mathbf{y}_{S_2}'[t+1]$. Then, $S_2$ subtracts $h_{RS_2}\sqrt{P_R}\tilde{\mathbf{x}}_{R}^{\prime(r)}[t-1]+h_{S_1S_2}\sqrt{P_{S_1}}\left(\tilde{\mathbf{x}}_{S_1}^{(r)}[t]+(1,\ldots,1)\right)$
 from the original received signal at $t^{th}$ block $\mathbf{y}_{S_2}'[t]$. The simplified form of this subtracted signal is
 \begin{equation}\label{y_S2_final}
   \mathbf{y}_{S_2}^{\prime\prime\prime}[t]=h_{S_1S_2}\sqrt{P_{S_1}}\mathbf{x}_{S_1}^{\prime (v)}[t]+\mathbf{e}_{S_2}+\mathbf{z}_{S_2}[t],
 \end{equation}
where $\mathbf{e}_{S_2}$ is the decoding error
\begin{eqnarray}
  \mathbf{e}_{S_2} &=& h_{S_1S_2}\sqrt{P_{S_1}}\left[\mathbf{x}_{S_1}^{(r)}[t]-\tilde{\mathbf{x}}_{S_1}^{(r)}[t]\right] \nonumber\\
   &+& h_{RS_2}\sqrt{P_R}\left[\hat{\mathbf{x}}_R^{\prime(r)}[t-1]-\tilde{\mathbf{x}}_{R}^{\prime(r)}[t-1] \right].
\end{eqnarray}
Then, $S_2$ uses  LDPC lattice decoding to find
\begin{equation}\label{vistigial_twoway}
   \tilde{\mathbf{b}}_{S_1}^{\prime (v)}[t]=\mathbf{L}^{(r)}\circ\left\lfloor \mathcal{D}\left( \textrm{DEC}_{\textrm{LDPCL}}\left(\mathbf{y}_{S_2},\boldsymbol{\sigma}_{S_2}\right) \right) \mathbf{G}_{\Lambda}^{-1}\right\rceil,
\end{equation}
where $\circ$ denotes the Hadamard product or entrywise product of matrices,  $y_{S_2,i}=\frac{y_{S_2,i}^{\prime\prime\prime}}{L_i^{(r)}h_{S_1S_2}\sqrt{P_{S_1}}}$ and  $\sigma_{S_2,i}=\frac{N_{S_2}}{L_i^{(r)}h_{S_1S_2}\sqrt{P_{S_1}}}$.

To obtain the vestigial information, $S_2$ computes $\tilde{\mathbf{b}}_{S_1}^{(v)}[t]=\mathrm{smod}(\tilde{\mathbf{b}}_{S_1}^{\prime (v)}[t],\mathbf{L}_{S_1})$. Finally, $S_2$ obtains the shaped and unshaped full-rate information  by taking the sum of the resolution and vestigial information as $\tilde{\mathbf{b}}_{S_1}'[t]=\tilde{\mathbf{b}}_{S_1}^{(r)}[t]+\tilde{\mathbf{b}}_{S_1}^{\prime(v)}[t]$ and $\tilde{\mathbf{b}}_{S_1}[t]=\tilde{\mathbf{b}}_{S_1}^{(r)}[t]+\tilde{\mathbf{b}}_{S_1}^{(v)}[t]$, respectively. A symbol error occurs at $S_2$ if $\tilde{b}_{S_1,i}\neq b_{S_1,i}$.

Note that the above decoding process is presented for the case that the employed shaping method is hypercube shaping. When the employed shaping is nested lattice shaping, this decoding steps are still valid by changing $\mathrm{smod}$ function into regular modulo operation. This is due to the fact that, the components of  lattice vectors, given as the inputs for hypercube shaping and nested lattice shaping, are drawn from different sets. For hypercube and nested lattice shading methods we use the sets $\left\{-L/2,\ldots,L/2-1\right\}$ and $\left\{0,\ldots,L-1\right\}$, respectively.
\section{Numerical Results}\label{Numerical Results}
\subsection{One-Way Relay Channels}
In the simulations, $|\mathcal{X}|=n/2$, i.e., we assume $50\%$ of the information integers are zero for resolution and vestigial information vectors.
We have used binary LDPC codes with $(n,k)=(1000,850),(5000,4250)$, where $n$ and $k$ are the codeword length and the dimension of the code, respectively. Symbol error rate (SER) performance of LDPC lattice codes are plotted against the sum power at source and the relay, i.e., $P_S E\left\{x_S^2 \right\}+P_R E\left\{x_R^2 \right\}$. We have considered $d_1=d_{SR} = 0.9$, $d_2=d_{RD} = 0.1$ and $d_{SD}=1$. The path loss exponents are $\alpha_1=1$, $\alpha_2 = 2$ . The variances of the noise at the relay and destination are $N_r = N_d = 0\mathrm{dB}$. The maximum number of iterations in each step of the decoding is assumed to be $50$.
Since, the encoder and the decoder both know the locations of the zeros in resolution and vestigial information, based on (\ref{si2})--(\ref{bvi}), for following locations we have
\begin{equation}\label{perdict}
 \left\{
\begin{array}{ll}
  x_{v,i}'=2b_{v,i}'-1=-1, & i\in \mathcal{I}^c\cap \mathcal{X}\\
  x_{r,i}'= 2b_{r,i}'-1=-1, & i\in \mathcal{I}^c\cap \mathcal{X}^c.
\end{array}
\right.
\end{equation}

We  estimate $E\left\{x_S^2\right\}$ and $E\left\{x_R^2\right\}$ for our scheme, for the case that hypercube shaping is applied. When, all of the elements of the lattice codewords are uniformly distributed over $(−L/2,L/2)$, the average power of $x_i$ is $E\left\{x_i^2\right\}=L_i^2/12$. Due to the fact that,  the resolution lattice vectors contain more zeros, the average power in this case is less than $L_i^2/12$.
We assume that the members of incoming integer  vector are uniformly distributed over $\mathcal{L}_i=\left\{-L_i/2,\ldots,L_i/2-1\right\}$, for $i=1,\ldots ,n$.
Since $E\left\{x_i\right\}=-0.5$ for $i\in \mathcal{I}^c$ and $E\left\{x_i\right\}=0$ for $i\in \mathcal{I}$, we have $E\left\{(2x_i-1)^2\right\}= 4E\left\{x_i^2\right\}+1$ for $i\in \mathcal{I}$, and $E\left\{(2x_i-1)^2\right\}= 4E\left\{x_i^2\right\}+3$ for $i\in \mathcal{I}^c$, and from above equations we have
\begin{eqnarray}
E\left\{x_i'^2\right\}=\left\{
\begin{array}{ll}
  4\times\frac{L_i^2+(L_i-2)^2}{24}+3, & i\in \mathcal{I}^c\\
  4\times \frac{4L_i^2}{12}+1, & i\in \mathcal{I}.
\end{array}
\right.
\end{eqnarray}
Put $L_1=L_2=\ldots =L_n=L$, then we have
\begin{eqnarray}
  E\left\{\mathbf{x}_S^2\right\} &=& \frac{\sum_{i=1}^n E\left\{x_i'^2\right\}}{n} \\
&=&\frac{k(L^2+(L-2)^2+18)+8(n-k)(L^2+6)}{6n}. \nonumber
\end{eqnarray}
For $i\in \mathcal{I}^c\cap \mathcal{X}^c$, $x_{r,i}'=-1$, so $E\left\{x_{r_i}'^2\right\}=1$. For  $i\in \mathcal{I}^c\cap \mathcal{X}$, $x_{r,i}'=2b_i-1$. Thus,
\begin{eqnarray}
  E\left\{\mathbf{x}_R^2\right\} &=& \frac{|\mathcal{I}^c\cap \mathcal{X}|\left((1/6)(L^2+(L-2)^2)+3\right)}{n}\nonumber \\
 &+&\frac{(n-k)(4L^2+1)}{3n}+\frac{|\mathcal{I}^c\cap \mathcal{X}^c|}{n}.
\end{eqnarray}

We have considered $L_i=8$, for $i=1,\ldots ,n$. Thus, for the cases that we have employed hypercube shaping, based on (\ref{rate}), the corresponding rate is $3.01$ bits/integer. For employing nested lattice shaping, we consider $L_1=\cdots =L_k=8$ and $L_{k+1}=\cdots =L_n=4$. Then, based on (\ref{rate_nested}), the corresponding rate is $2.85$ bits/integer.
In order to achieve these rates, according to (\ref{rate_bound}), the total required  powers  are $P_1=P_S E\left\{\mathbf{x}_S^2 \right\}+P_R E\left\{\mathbf{x}_R^2 \right\}\geq 51.88= 17.15$dB, and $P_2\geq 41.2= 16.15$dB, respectively.

\begin{figure}[!h]
\centering
\includegraphics[width=4in]{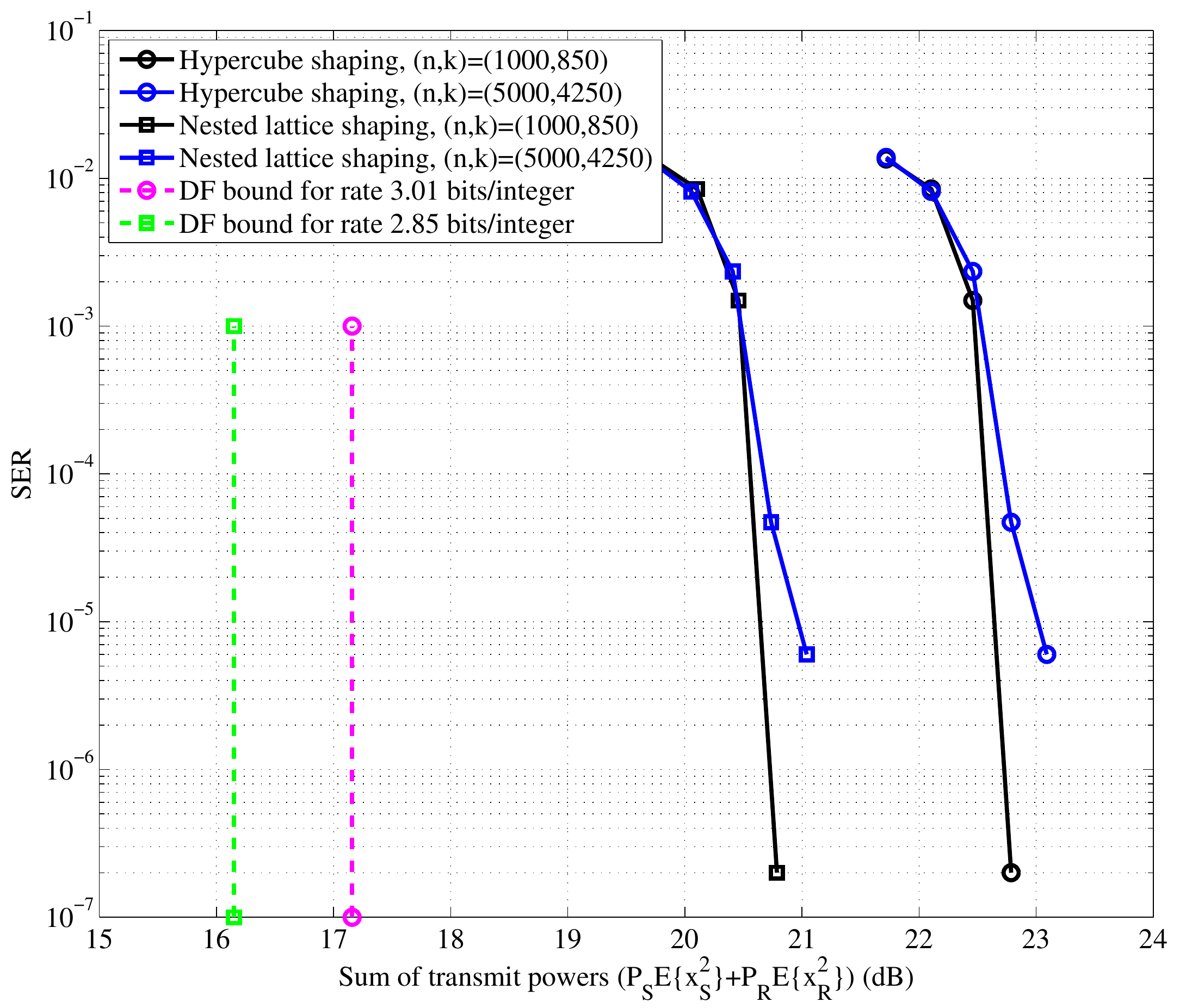}
\caption{Symbol error rate of LDPC lattice codes over the one-way relay channel.}
\label{figsim1}
\end{figure}

In \figurename{\ref{figsim1}}, we have presented SER variation versus sum of transmit powers for both nested-lattice shaping and hypercube shaping. In~\cite{DBLP:conf/icc/FerdinandNA13}, the implementation of block Markov encoding proposed for LDLC lattice codes. We have considered $h_{SD}$, $h_{SR}$, $h_{RD}$ and other parameters similar to their corresponding values in~\cite{DBLP:conf/icc/FerdinandNA13}.  The SER performance of an LDLC lattice code with dimension $1000$ and rate $2.78$, which is obtained by employing nested lattice shaping, at $10^{-4}$ is $3.77\textrm{dB}$ away from its corresponding DF inner bound, which is $15.77$dB. We observe that the SER performance of an LDPC lattice code of length $1000$  at $10^{-4}$ is $4.5\textrm{dB}$ away from its corresponding DF inner bound. This is a natural result, due to the better SER performance of LDLCs comparing to LDPC lattice codes, over AWGN channels. Different decoders have been proposed for LDLCs. As far as we know, the best one is proposed in \cite{5205636}.
The decoding complexity of LDLCs, by using proposed decoder in \cite{5205636},  is at least $24$ times more than the decoding complexity of  LDPC lattices. Indeed, the decoding complexity of an  LDPC lattice of dimension $1000$ is equivalent to the decoding complexity of an LDLC with dimension $24000$. Results of \figurename{\ref{figsim1}} show that the increase in the dimension of the lattice can decrease the gap between DF bound and the performance curve. Using an LDPC lattice code of dimension $5000$ instead of dimension $1000$  makes about $0.55$dB improvement in the performance.
\subsection{Two-Way Relay Channels}
\begin{figure}[!h]
\centering
\includegraphics[width=4in]{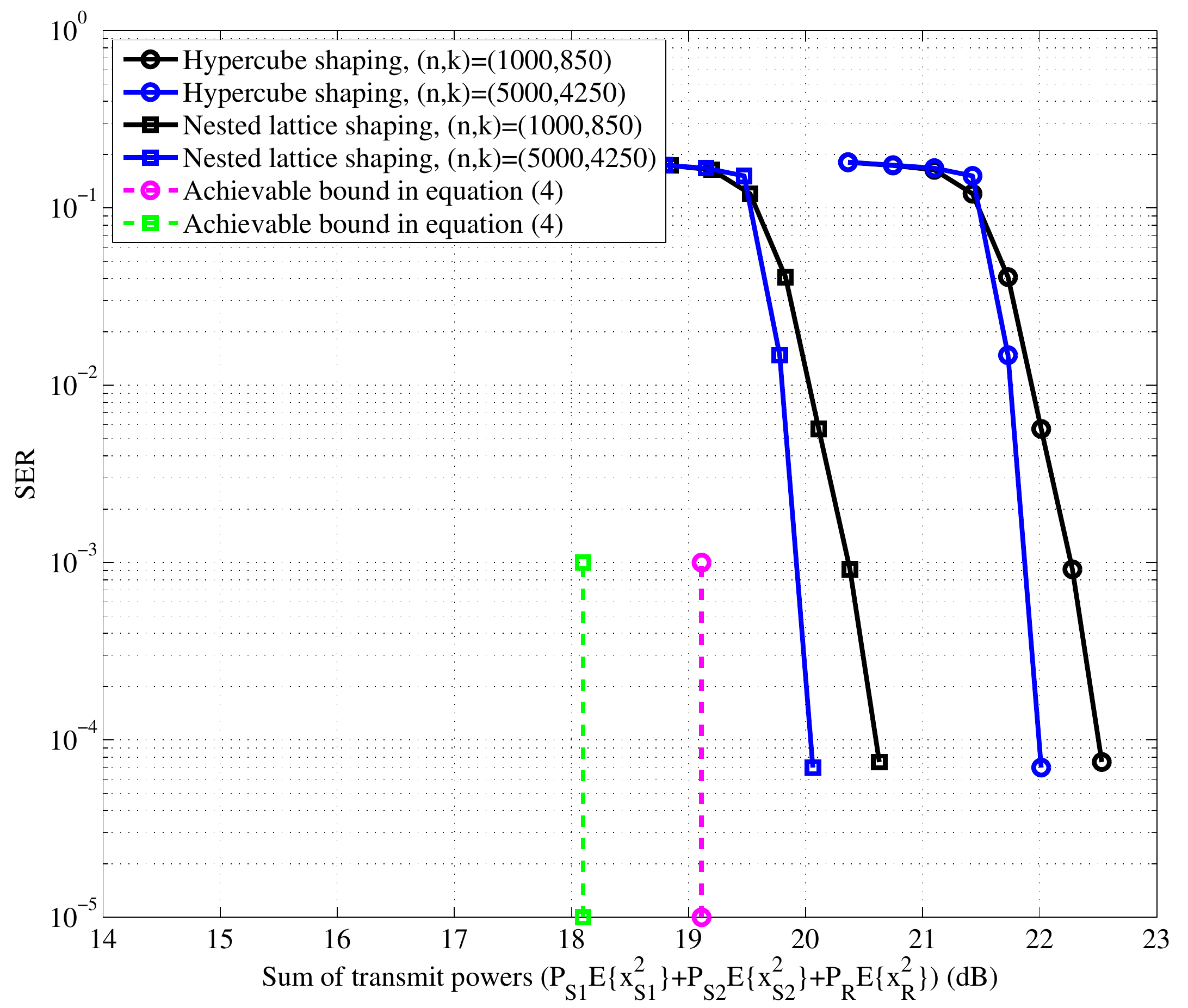}
\caption{Symbol error rate of  LDPC lattice codes over the two-way relay channel. }
\label{figsim2}
\end{figure}

In \figurename{\ref{figsim2}}, we plot the SER versus the sum of transmit powers $\left(\sum_{i=1}^2P_{S_i}E\{\mathbf{x}_{S_i}^2\}+P_{R}E\{\mathbf{x}_{R}^2\}\right)$, for the two-way relay channel. We suppose that the relay is midway between the sources, that is, $d_{S_1R}=d_{S_2R}=0.5$ and $d_{S_1S_2}=1$. We assume $L_{S_1,i}=L_{S_2,i}=8$, for $i=1,\ldots,k$ and $L_{S_1,i}=L_{S_2,i}=4$, for $i=k+1,\ldots,n$. For the resolution lattice we consider $\mathbf{L}^{(r)}=(2,\ldots,2)$. We choose $N_{S_1}=N_{S_2}=N_{R}=0$dB. Path-loss exponents
are $\alpha_{S_1R}=\alpha_{RS_1}=1$ and $\alpha_{RS_2}=\alpha_{S_2R}=5$. The used underlying codes are the same ones that we used for one-way relay channels.
Depending on the dimension of the lattice and  the applied shaping method, our scheme achieves to within $2$dB of the achievable rate in (\ref{rate_bound2}).
\section{Conclusions}\label{conclusions}
In this paper, we present the implementation of block Markov encoding using LDPC lattice codes over the one-way and two-way relay channels. In order to apply this scheme, we employ a low complexity decoding method for LDPC lattices. Then, for using these lattices in the power constrained scenarios, we propose two efficient shaping methods based on hypercube shaping and nested lattice shaping. We apply different decomposition schemes for one-way
and two-way relay channels. The applied decomposition schemes
are the altered versions of the applied methods for decomposing
LDLCs. Due to the lower complexity of decoding LDPC lattices comparing to LDLCs, the
complexity of the proposed schemes in this paper are significantly
lower than the ones proposed for LDLCs. Simulation results show that LDLCs outperform  LDPC lattices in general. However, having lower decoding complexity enables us to increase the dimension of the lattice to  compensate the existing gap between the performance of the  LDPC lattice codes and the LDLCs.
\end{document}